\documentclass[pra,a4paper,twocolumn,superscriptaddress,longbibliography,nofootinbib]{revtex4-2}
\usepackage[utf8]{inputenc}
\usepackage{amssymb}
\usepackage{amsmath}
\usepackage{amsfonts}
\usepackage{amsthm}
\usepackage{amsbsy}
\usepackage{bm}
\usepackage{bbold}

\usepackage{graphicx}
\usepackage{xcolor}
\usepackage{multirow}
\usepackage{float}
\usepackage{comment}
\usepackage{ulem}
\usepackage{relsize}
\usepackage{cmap}
\usepackage{physics}
\usepackage[colorlinks]{hyperref}

\DeclareMathOperator{\sgn}{sgn}

\DeclareMathOperator{\so}{\textsf{s}_\textsf{0}}
\DeclareMathOperator{\sx}{\textsf{s}_\textsf{1}}
\DeclareMathOperator{\sy}{\textsf{s}_\textsf{2}}
\DeclareMathOperator{\sz}{\textsf{s}_\textsf{3}}
\DeclareMathOperator{\sj}{\textsf{s}_\textsf{j}}
\DeclareMathOperator{\sk}{\textsf{s}_\textsf{k}}

\DeclareMathOperator{\li2}{li_2}
\DeclareMathOperator{\llangle}{\langle\!\langle}
\DeclareMathOperator{\rrangle}{\rangle\!\rangle}
\DeclareMathOperator{\Llangle}{\Bigl \langle\!\!\Bigl \langle}
\DeclareMathOperator{\Rrangle}{\Bigr \rangle\!\!\Bigr\rangle}
\DeclareMathOperator{\DD}{\mathcal{D}\!\!\mathcal{D}}
\DeclareMathOperator{\TT}{\texttt{T}}
\DeclareMathOperator{\TP}{\texttt{P}}

\newcommand{\DOS}{{LDoS}}
\newcommand{\NLSM}{{NL$\sigma$M}}

\begin{document}
	
	\title{Generalized multifractality in the spin quantum Hall symmetry class with interaction}
	
	\author{S. S. Babkin}
	
	\affiliation{Moscow Institute of Physics and Technology, 141700 Dolgoprudnyi, Moscow Region, Russia}
	\affiliation{\hbox{L.~D.~Landau Institute for Theoretical Physics, acad. Semenova av. 1-a, 142432 Chernogolovka, Russia}}

	\author{I. S. Burmistrov}
	
	\affiliation{\hbox{L.~D.~Landau Institute for Theoretical Physics, acad. Semenova av. 1-a, 142432 Chernogolovka, Russia}}

	\date{\today} 
	
\begin{abstract}
Scaling of various local observables with a system size at Anderson transition criticality is characterized by a generalized multifractality. We study the generalized multifractality in the spin quantum Hall symmetry class (class C) in the presence of interaction. We employ Finkel'stein nonlinear sigma model and construct the pure scaling derivativeless operators for class C in the presence of interaction. Within the two-loop renormalization group analysis we compute the anomalous dimensions of the pure scaling operators and demonstrate that they are affected by the interaction. We find that the interaction breaks 
exact symmetry relations between generalized multifractal exponents known for a noninteracting problem.

	\end{abstract}

	\maketitle
	
\section{Introduction}
	
Anderson transition is  a fascinating example of a disorder-driven quantum phase transition separating metallic and insulating phases. A nontrivial topology, e.g. as in the integer quantum Hall effect, makes Anderson transition to occur between distinct topological phases. Although more than 60 years past from the seminal paper by Anderson \cite{Anderson58}, localization--delocalization transitions in disordered media are still a subject of intense research. A striking feature of Anderson transition is strong mesoscopic fluctuations of electron wave functions or, equivalently, the local density of states ({\DOS}) \cite{Wegner1980,Castellani1986,Lerner1988}. At criticality the disorder averaged moments of {\DOS} demonstrate {\it pure} power-law scaling with the system size, $\langle \rho^q\rangle \sim L^{-x_{\textsf{(q)}}}$, with the {\it multifractal} exponents $x_{\textsf{(q)}}$ whose values depend on a symmetry class (see Refs. \cite{Mirlin2000,EversMirlin} for a review). 

In fact, there are much more observables than just the moments of {\DOS} that demonstrate pure scaling with $L$ at critically \cite{Wegner1986}. Some time ago, the corresponding observables were constructed in terms of specific correlations of wave functions for the case of unitary Wigner--Dyson class (class A) \cite{Gruzberg2013}. Moreover, it was proven that the corresponding multifractal exponents $x_\lambda$ are not unrelated but satisfy a set of symmetry relations specific for each symmetry class \cite{Mirlin2006,Gruzberg2011,Gruzberg2013}. Therefore, the set of multifractal exponents $x_\lambda$, termed as {\it generalized multifractality}, is a unique fingerprint of a certain Anderson transition.   

Although a nonlinear sigma model ({\NLSM}) description for each of ten Altland--Zirnbauer symmetry classes is known (see Ref. \cite{EversMirlin} for a review), ultimate theories for Anderson transition criticality are still to be found. A well-known example of long-time search for such critical theory is  the integer quantum Hall plateau transition. In Refs. \cite{Zirnbauer1999,Kettemann1999,Bhaseen2000,Tsvelik2001,Tsvelik2007,Zirnbauer2019} the Wess--Zumino--Novikov--Witten models were conjectured as theories for quantum Hall criticality. These theories predict  multifractal exponents $x_{\textsf{(q)}}$ to be quadratic functions of $q$. In fact, the parabolic multifractal spectrum is more general. It can be obtained under assumptions of the local conformal invariance and Abelian fusion rules \cite{Bondesan2017}. However, present numerical computations of multifractal spectrum demonstrate significant deviations from the exact parabolicity \cite{Obuse2008,Evers2008}. This makes the theoretical predictions of parabolic multifractal spectrum for the integer quantum Hall criticality to be questionable. 

Recently,  validity of the local conformal invariance has been tested in superconducting cousin of the integer quantum Hall effect --- the spin quantum Hall effect (class C) in two-dimensions \cite{Kagolovsky1999,Senthil1999}.  It was shown \cite{Karcher2021} that the assumption of the local conformal invariance
at the spin quantum Hall transition leads to the parabolicity of the generalized multifractal spectrum $x_\lambda$. An advantage of the spin quantum Hall transition in two dimensions, $d=2$, is that a subset of multifractal exponents can be found exactly by mapping to the percolation problem \cite{Gruzberg1999,Beamond2002,Mirlin2003,Evers2003,Subramaniam2008,Karcher2022}. These exact analytical values of exponents serve as a benchmark against numerical computations. Although the numerical simulations reproduce exact analytical results, they demonstrate clear evidence for a violation of parabolicity of the generalized multifractal spectrum \cite{Mirlin2003,Puschmann2021,Karcher2021}. 
These results signal a lack of the local conformal invariance at the spin quantum Hall transition in $d=2$.

Multifractality is not only an interesting playground for theoretical and numerical analysis. A signature of multifractal behavior has been found experimentally in light waves spreading in an array of dielectric nanoneedles \cite{Mascheck2012} as well as 
 in ultrasound waves propagating through a system of randomly packed Al beads \cite{Faez2009}. Also the multifractality of an electron {\DOS}  has been reported in diluted magnetic semiconductor Ga$_{1{-}x}$Mn$_x$As \cite{Richardella}. 

Multifractal correlations of wave functions result in a variety of interesting physical effects. In particular, they lead to strong enhancement of superconducting transition temperature and the superconducting gap at zero temperature \cite{FeigelmanYuzbashyan2007,FeigelmanCuevas2010,Burmistrov2012,Burmistrov2015b,DellAnna,Burmistrov2021}, induce log-normal distribution for the superconducting order parameter \cite{FeigelmanCuevas2010,Fan2020,Stosiek2020} and local density of states \cite{Burmistrov2021,Stosiek2021} in dirty superconductors, are responsible for instabilities of surface states in topological superconductors \cite{Foster2012,Foster2014}, result in strong mesoscopic fluctuations of the Kondo temperature \cite{Kettemann2006,Micklitz2006,Kettemann2007}, enhance depairing effect of magnetic impurities on superconducting state in dirty films \cite{Burmistrov2018}, make cooling of electrons due to electron-phonon coupling more efficient \cite{Feigelman2019}, and affect the Anderson orthogonality catastrophe \cite{Kettemann2016}.

The scaling properties at Anderson transition criticality can be modified by electron-electron interaction if it is a relevant perturbation. In this case the so-called Mott--Anderson transition emerges being controlled by both disorder and interaction strengths (see Refs. \cite{Fin,Belitz1994} for a review). Surprisingly, as it has been shown recently, not only the multifractality 
\cite{Burmistrov2013,Burmistrov2015m}
but also the generalized multifractality \cite{Repin2016} 
exists at Mott-Anderson criticality in standard Wigner--Dyson symmetry classes.  
In this case the pure scaling operators correspond to proper correlations of single particle Green's functions. An important example is the moments of the {\DOS} that in the presence of interaction remain pure scaling operators \cite{Burmistrov2013,Burmistrov2015m}. Interaction drives  a disordered electron system into a new interacting fixed point corresponding to a metal-insulator transition and, consequently, affects the generalized multifractal exponents. Interestingly, the symmetry relations between the multifractal exponents in standard Wigner--Dyson symmetry classes survive in the presence of interaction, at least within the second order expansion in $\epsilon=d-2$. Unfortunately, present numerical computing power is not enough to access generalized multifractal exponents and to check symmetry relations in the presence of interaction \cite{Slevin2012,Slevin2014,Amini2014,Lee2018}.

In this paper we develop the theory of the generalized multifractality for the spin quantum Hall symmetry class in the presence of electron-electron interaction. Using Finkel'stein {\NLSM} for class C, we demonstrate that the pure scaling derivativeless operators can be constructed by straightforward generalization of the pure scaling operators without derivatives known in the absence of interaction. Within the two-loop approximation we compute how the anomalous dimensions of the pure scaling derivativeless operators are affected by the presence of interaction. Applying our results to the transition in $d=2+\epsilon$ dimensions, we illustrate breakdown of the exact symmetry relations between generalized multifractal exponents for class C in the presence of interaction, cf. Eq. \eqref{eq:xLambda:2+e}. Also for a reader's convenience, within the Finkel'stein {\NLSM} we rederive the results known in the literature  for the one-loop renormalization of the spin conductance, dimensionless interaction, the Finkel'stein frequency renormalization parameter, and the averaged {\DOS}.

The outline of the paper is as follows. In Sec. \ref{Sec:Formalism} we introduce formalism of Finkel'stein {\NLSM} for class C. The details for background field renormalization of the action is presented in Sec. \ref{sec:BGFR}. The linear response is studied in Sec. \ref{Sec:LinearR}. In Sec. \ref{Sec:Operators} local derivativeless operators are constructed and their anomalous dimensions are computed within the two-loop approximation.  We end the paper with discussions and conclusions in Sec. \ref{Sec:Final}.

\section{Formalism of Finkel'stein {\NLSM} \label{Sec:Formalism}}
	
\subsection{{\NLSM} action}	
	
We start with the description of formalism of the Finkel'stein {\NLSM} applied to class C. We follow approach of Refs. \cite{Bruno2005,DellAnna2006}. As usual, the {\NLSM} action is given as a sum of the noninteracting part, $S_{\rm 0}$, and the term $S_{\rm int}$, describing interaction: 
\begin{equation}
Z=\int D[Q] \exp S, \qquad S=S_{\rm 0} + S_{\rm int},
\label{eq:NLSM}
\end{equation}
where
\begin{subequations}
\begin{align}
S_{\rm 0} & = -\frac{g}{16} \int_{\bm{x}} \Tr (\nabla Q)^2 + Z_\omega \int_{\bm{x}} \Tr \hat\varepsilon  Q ,
\label{eq:S:0-0}
 \\
S_{\rm int}& =-\frac{\pi T \Gamma_t}{4} \sum_{\alpha,n} 
\int_{\bm{x}} \Tr (I_n^\alpha \bm{\textsf{s}} Q)
 \Tr (I_{-n}^\alpha \bm{\textsf{s}} Q) .
 \label{eq:S:int}
\end{align}
\end{subequations}
Here and in what follows, we use the shorthand notation $\int_{\bm{x}}\equiv \int d^d\bm{x}$. The field $Q$ is Hermitian matrix, $Q^\dag=Q$, which satisfies a standard nonlinear local constraint
\begin{equation}
Q^2(\bm{x})=1 .
\label{eq:nonlin-const}
\end{equation}   
The matrix field $Q$  acts in the $N_r\times N_r$ replica space, in the $2\times 2$ spin space and  in the $2N_m\times 2N_m$ space of the Matsubara fermionic energies, $\varepsilon_n=\pi T(2n+1)$. 
The action \eqref{eq:NLSM} involves the following matrices\begin{gather}
(I_k^\gamma)_{nm}^{\alpha\beta}=\delta_{n-m,k}\delta^{\alpha\beta}\delta^{\alpha\gamma} \so
,\quad 
\hat \varepsilon_{nm}^{\alpha\beta}=\varepsilon_n \, \delta_{nm}\delta^{\alpha\beta} \so .
\end{gather}
Here $\so$ denotes the $2\times 2$ identity matrix in the spin space.
We note that Greek indices represent replica space whereas Latin indices corresponds to Matsubara energies. The vector $\bm{\textsf{s}}=\{\sx,\sy,\sz\}$ is the vector of three nontrivial Pauli matrices
\begin{equation}
\sx = \begin{pmatrix}
0 & 1\\
1 & 0
\end{pmatrix}, \quad
 \sy= \begin{pmatrix}
0 & -i\\
i & 0
\end{pmatrix},  \quad 
\sz = \begin{pmatrix}
1 & 0\\
0 & -1
\end{pmatrix} .
\end{equation}

Since spin quantum Hall symmetry class (class C) belongs to the Bogolubov -- de Gennes symmetry classes there is 
an additional symmetry that relates positive and negative Matsubara energies,
\begin{equation}
\begin{split}
Q= & - \bar{Q}, \qquad \bar{Q}= \sy L_0 Q^\textsf{T} L_0 \sy ,  \\
& (L_0)_{nm}^{\alpha\beta}=\delta_{\varepsilon_n,-\varepsilon_m}\delta^{\alpha\beta} \so .
\end{split}
\label{eq:symm:C}
\end{equation}
Here superscript $^\textsf{T}$ denotes the matrix transposition operation.

Nonlinear constraint \eqref{eq:nonlin-const} can be resolved by representing the matrix $Q$ as rotation around the fixed matrix $\Lambda$: 
\begin{equation}
Q = \texttt{T}^{-1}\Lambda \texttt{T}, \qquad  \Lambda_{nm}^{\alpha\beta} = \sgn \varepsilon_n \, \delta_{nm} \delta^{\alpha\beta}\so
\label{eq:T-rep}
\end{equation}
Here the rotation $\texttt{T}$ is a unitary matrix satisfying 
\begin{equation}
\texttt{T}^{-1}=\texttt{T}^\dag, \quad  (\texttt{T}^{-1})^\textsf{T} L_0 \sy= \sy L_0 \texttt{T} .
\label{eq:relations:T}
\end{equation}

The {\NLSM} action \eqref{eq:NLSM} involves three parameters. Bare dimensionless spin conductance is denoted as $g$. Bare strength of exchange interaction (interaction in the triplet particle-hole channel) is $\Gamma_t$. The parameter $Z_\omega$ describes the renormalization of the frequency term. Generically, $g$, $\Gamma_t$, and $Z_\omega$ are subjected to renormalization. Finally, temperature is denoted by $T$.

The constraint \eqref{eq:symm:C} and the parametrization \eqref{eq:T-rep} determines the target space of the {\NLSM} as $Q\in G/K=\textrm{Sp}(2N)/\textrm{U}(N)$ where $N=2N_rN_m$. Indeed
the Hermitian matrix satisfying constraint \eqref{eq:symm:C} can be parametrized by $2N^2+N$ real variables. This corresponds to $G=\textrm{Sp}(2N)$. The nonlinear 
constraint \eqref{eq:nonlin-const}, or, equivalently, the parametrization \eqref{eq:T-rep} with a unitary matrix $\texttt{T}\in \textrm{U}(N)$ (since it obeys the relation \eqref{eq:relations:T}), reduces the number of real variables down to $N^2+N$. The latter corresponds to the symmetric space  $G/K=\textrm{Sp}(2N)/\textrm{U}(N)$.

We note that the symmetry \eqref{eq:symm:C} forbids the interaction in singlet particle-hole channel since $\Tr I_n^\alpha \so Q \equiv 0$. The Cooper channel interaction is suppressed by the absence of time-reversal symmetry.

The action \eqref{eq:NLSM} of the Finkel'stein {\NLSM} for the class C 
is similar to the one for a standard Wigner-Dyson class A in the presence of spin rotation symmetry (see Refs. \cite{Fin,Belitz1994,Burmistrov2019} for review). We emphasize two distinctions. At first, there is no interaction in the singlet particle-hole channel. Secondly, there is the additional symmetry relation \eqref{eq:symm:C}. These two features make class C in the presence of interaction to be different from interacting class A. 

Similarly to the class A, the action \eqref{eq:NLSM} can be supplemented by the Pruisken's theta-term. Being topological this term does not affect perturbative analysis presented in this paper however it is responsible for the existence of the spin quantum Hall transition in $d=2$ dimensions. 

\subsection{Perturbation theory}

In order to construct perturbation theory, we prefer to use the square-root parametrization of the $Q$ matrix:
\begin{equation}
Q = W+\Lambda \sqrt{1-W^2}, \qquad W = \begin{pmatrix} 
0 & w \\
w^\dag & 0
\end{pmatrix} .
\end{equation}
Here the block structure of matrix $W$ is with respect to positive and negative Matsubara frequencies. In particular, we adopt the following notations: $W_{n_1n_2} = w_{n_1n_2}$ and $W_{n_2n_1} = w^\dag_{n_2n_1}$ with $\varepsilon_{n_1}> 0$ and $\varepsilon_{n_2}<0$.
It is convenient to use the following expansion $w=\sum_{\textsf{j}=0}^3 w_\textsf{j} \sj$.
As a consequence of the constraint \eqref{eq:symm:C}, the elements of $w_\textsf{j}$
satisfy the symmetry relations
\begin{equation}
\begin{split}
 & (w_\textsf{j})^{\alpha\beta}_{n_1n_2}  = \textsf{v}_\textsf{j}  (w_\textsf{j})^{\beta\alpha}_{-n_2,-n_1},  \\
& \textsf{v}_\textsf{j} = -\tr (\sj \sy \textsf{s}_\textsf{j}^\textsf{T} \sy)/2 = \{-1,1,1,1\} .
\end{split}
\label{eq:const:C}
\end{equation}
In particular, Eq. \eqref{eq:const:C} implies $(w_\textsf{0})^{\alpha\alpha}_{n_1,-n_1}\equiv 0$. 
	
Expanding the action \eqref{eq:NLSM} to the second order in $W$, we find the propagators of Gaussian theory:	
\begin{gather}
{}\hspace{-0.5cm}\Bigl  \langle (w_\textsf{j})^{\alpha\beta}_{n_1n_2}(\bm{q})  (w_\textsf{j}^\dag)^{\mu\nu}_{n_4n_3}(-\bm{q}) \Bigr \rangle 
 = \frac{2}{g} 
 \Bigl [\delta^{\alpha\nu}\delta^{\beta\mu}\delta_{n_1n_3}\delta_{n_2n_4}
\notag \\
+ \textsf{v}_\textsf{j}\delta^{\alpha\mu}\delta^{\beta\nu}\delta_{n_1,-n_4}\delta_{n_2,-n_3}  
- \frac{4\pi T\gamma}{D}(1-\delta_{\textsf{j}0})\delta^{\alpha\nu}\delta^{\beta\mu}\delta^{\alpha\beta}
\notag \\
\times
\delta_{n_{12},n_{34}}\mathcal{D}^t_q(i\omega_{n_{12}}) \Bigr ] 
\mathcal{D}_q(i\omega_{n_{12}}) .
\label{eq:prop:full}
\end{gather}
Here we use the following shorthand notations $n_{12}=n_1-n_2$ and $\omega_{n_{12}}=\varepsilon_{n_1}-\varepsilon_{n_2}$. Also we introduced the bare diffusion coefficient $D=g/(4Z_\omega)$ and dimensionless interaction strength $\gamma=\Gamma_t/Z_\omega$. Next, 
\begin{subequations}
\begin{align}
\mathcal{D}_q(i\omega_n) & =\Bigl [ q^2+\omega_n/D\Bigr ]^{-1}, \label{eq:prop:def:a}\\
\mathcal{D}^t_q(i\omega_n) & =\Bigl [ q^2+(1+\gamma)\omega_n/D\Bigr ]^{-1}
\label{eq:prop:def:b}
\end{align}
\end{subequations}
\noindent{stand} for diffuson and diffuson dressed by interaction via ladder resummation, respectively. We mention that the product $\gamma  \mathcal{D}^{-1} _q(i\omega_n)\mathcal{D}^t_q(i\omega_n)$ gives the dynamically screened exchange interaction in the random phase approximation.

As we shall see below, in the process of renormalization of the {\NLSM} action it is convenient not to keep track on Matsubara frequencies of slow fields in the propagators. Then, in order to regularize the infrared, it is convenient to add the following regulator into the action \eqref{eq:NLSM}:
 \begin{equation}
S_{\rm h} = \frac{g h^2}{8} \int_{\bm{x}} \Tr \Lambda Q .
\label{eq:S:reg:h}
\end{equation}
On the level of the Gaussian theory $S_{\rm h}$ results in the change $q^2\to q^2+h^2$ in the diffusive propagators \eqref{eq:prop:def:a} and \eqref{eq:prop:def:b}.

\section{Background field renormalization of the action\label{sec:BGFR}}

In this section we present details of the one-loop renormalization of  the {\NLSM} action within the background field method. Although one-loop results for parameters $g$, $Z_\omega$, and $\gamma$ has been reported before in Refs. \cite{Jeng2001a,Jeng2001,DellAnna2006,Liao2017}, 
 this section serves place to set notations. 

Let us split the matrix field: $Q \to \texttt{T}^{-1} Q \TT$ where $Q$ now plays the role of ``fast'' mode and $\underline{Q}=\texttt{T}^{-1}\Lambda \TT$ is  a ``slow'' field. We assume that the matrix field $\TT$ deviates from the unit matrix only at small frequencies such that 
$\TT_{nm}^{\alpha\beta} = \delta_{nm}\delta^{\alpha\beta}\so$ for $\max\{|\varepsilon_n|,|\varepsilon_m|\}\gg \Omega$. Here $\Omega$ plays the role of the ultra-violet cutoff for ``slow'' modes. Next we write
\begin{equation}
S[\texttt{T}^{-1} Q \TT] = S[\underline{Q}]+S[Q] + \delta S_{\rm 0}+\delta S_{\rm int} ,
\end{equation}
where  
\begin{subequations}
\begin{align}
 \delta S_{\rm 0} & = \delta S_{\rm 0}^{(1)}+\delta S_{\rm 0}^{(2),1}
 +\delta S_{\rm 0}^{(2),2} +\delta S_{\rm 0}^{(\varepsilon)}+\delta S_{\rm 0}^{\rm (h)} , \\
\delta S_{\rm int} &  = \delta S_{\rm int}^{(1),1}+\delta S_{\rm int}^{(1),2}+\delta S_{\rm int}^{(2),1}+\delta S_{\rm int}^{(2),2} . 
\end{align}
\end{subequations}
Here following Ref. \cite{Baranov1999b}, we introduce the following notations 
\begin{subequations}
\begin{align}
\delta S_{\rm 0}^{(1)} = & - \frac{g}{4} \int_{\bm{x}}\Tr \bm{\texttt{A}}\overline{\delta Q} \nabla \overline{\delta Q} , \\
\delta S_{\rm 0}^{(2),1} = & - \frac{g}{4} \int_{\bm{x}}\Tr \bm{\texttt{A}}\overline{\delta Q} \bm{\texttt{A}}\Lambda ,\\
\delta S_{\rm 0}^{(2),2} = & - \frac{g}{4} \int_{\bm{x}}\Tr \bm{\texttt{A}}\overline{\delta Q} \bm{\texttt{A}}\overline{\delta Q} ,\\
\delta S_{\rm 0}^{(\varepsilon)}= & Z_\omega  \int_{\bm{x}} \Tr [\TT \hat \varepsilon,\texttt{T}^{-1}] \overline{\delta Q},\\
\delta S_{\rm 0}^{(h)}= &\frac{g h^2}{8}  \int_{\bm{x}} \Tr [\TT \Lambda,\texttt{T}^{-1}] \overline{\delta Q}, \\
\delta S_{\rm int}^{(1),1}  = & - \frac{1}{2}\pi T\Gamma_t \sum_{\alpha, n} 
\int_{\bm{x}} \tr I^\alpha_n \bm{\textsf{s}} \underline{Q}   \tr I^\alpha_{-n} \bm{\textsf{s}} \overline{\delta Q},   \\
\delta S_{\rm int}^{(2),1}  = & - \frac{1}{2}\pi T\Gamma_t \sum_{\alpha, n} 
\int_{\bm{x}} \tr I^\alpha_n \bm{\textsf{s}} \underline{Q}   \tr \bm{\texttt{A}}^\alpha_{-n} \overline{\delta Q}  ,  \\
\delta S_{\rm int}^{(1),2}  = & - \frac{1}{2}\pi T\Gamma_t \sum_{\alpha, n} 
\int_{\bm{x}} \tr I^\alpha_{n} \bm{\textsf{s}} \overline{\delta Q}   \tr \bm{\texttt{A}}^\alpha_{-n}  \overline{\delta Q} ,  \\
\delta S_{\rm int}^{(2),2}  = & - \frac{1}{4}\pi T\Gamma_t \sum_{\alpha, n} 
\int_{\bm{x}} \tr \bm{\texttt{A}}^\alpha_{n}  \overline{\delta Q}  
 \tr \bm{\texttt{A}}^\alpha_{-n} \overline{\delta Q}  ,
\end{align}
\end{subequations}
where $\overline{\delta Q}= Q-\Lambda$, $\bm{\texttt{A}}=\TT\nabla \texttt{T}^{-1}$, and  $\bm{\texttt{A}}_n^\alpha = [\TT I^\alpha_{n} \bm{\textsf{s}},\texttt{T}^{-1}]$ .

Within the one-loop approximation, the effective action for the ``slow'' field $\underline{Q}$ can be obtain as
\begin{gather}
S_{\rm eff}[\underline{Q}] = \ln \int D[Q] e^{S[\texttt{T}^{-1} Q \TT]} 
\simeq S[\underline{Q}]+ \langle \delta S_{\rm 0}+\delta S_{\rm int} \rangle 
\notag \\
+ \frac{1}{2} \llangle(\delta S_{\rm 0}+\delta S_{\rm int})^2\rrangle .
\end{gather}  
Here $\llangle A^2\rrangle$ denotes the irreducible average.
We note that it is enough to expand $\overline{\delta Q}$ to the second order in $W$ for computation of the averages in the above expressions.

\subsection{Background field renormalization of $\Gamma_t$}

We start computation from renormalization of the exchange interaction $\Gamma_t$. There are several contributions. At first, we find
\begin{gather}
\langle \delta S_{\rm int}^{(2),1} \rangle \simeq 
\frac{1}{4}\pi T\Gamma_t \sum_{\alpha, n} 
\int_{\bm{x}} \tr I^\alpha_n \bm{\textsf{s}} \underline{Q} \tr \bm{\texttt{A}}^\alpha_{-n} \Lambda  \langle
W^2\rangle  \notag\\
\to
\frac{\delta_{(2),1}\Gamma_t}{\Gamma_t} S_{\rm int}[\underline{Q}] , 
\end{gather}
where 
\begin{gather}
\frac{\delta_{(2),1}\Gamma_t}{\Gamma_t} =  - \frac{2 \textsf{v}}{g} \int\limits_p\mathcal{D}_p(0) +
\frac{24\pi T \gamma}{g D} \sum_{m>0} \int\limits_p\DD_p^t(i\omega_m) .
\label{eq:Gammat:21}
\end{gather}
Here, in order to keep track for anomalous contribution related with the additional symmetry \eqref{eq:symm:C}, we introduce the parameter $\textsf{v}=\sum_{\textsf{j}=0}^3 \textsf{v}_\textsf{j} =2$. Also we introduce the shorthand notations which will be intensively used below: $\int_p \equiv \int d^d\bm{p}/(2\pi)^d$
and $\DD_p^t(i\omega_m)\equiv\mathcal{D}_p(i\omega_m)
\mathcal{D}_p^t(i\omega_m)$. 

Next contribution to $\Gamma_t$ comes from
\begin{gather}
\frac{1}{2} \llangle\! (\delta S_{\rm int}^{(1),1})^2 \! \rrangle  {\simeq} 
\frac{1}{2}\! \Llangle\!\! \left (\! \frac{\pi T\Gamma_t}{4}\!\!  \sum_{\alpha;n}
\!\int_x \!\tr I^\alpha_n \bm{\textsf{s}} \underline{Q}  
 \tr I^\alpha_{-n} \bm{\textsf{s}} \Lambda W^2 \right )^2\!\Rrangle
 \notag \\
 \to \frac{\delta_{(1),1;(1),1}\Gamma_t}{\Gamma_t} S_{\rm int}[\underline{Q}] 
 , 
 \end{gather}
 where
 \begin{align}
 \notag \\
\frac{\delta_{(1),1;(1),1}\Gamma_t}{\Gamma_t} & =  - \frac{8\pi T \gamma}{g D}\sum_{n>0}\int_p \mathcal{D}^2_p(2i\varepsilon_n) \notag \\
& {}\hspace{-1cm} +
\frac{16\pi T \gamma}{g D} \sum_{m>0} \int_p\Bigl [\mathcal{D}_p^2(i\omega_m)-
\mathcal{D}^{t2}_p(i\omega_m)
\Bigr ] .
\label{eq:Gammat:11:11}
\end{align}
One more contribution is as follows
\begin{gather}
\llangle\! \delta S_{\rm int}^{(1),1} \delta S_{\rm int}^{(1),2}\!\rrangle\! =\!
- \frac{\left (\pi T\Gamma_t\right )^2}{8}\! 
\Llangle\!\sum_{\alpha,n}\!\int\limits_{\bm{x}}\!\! \tr I^\alpha_n \bm{\texttt{s}} \underline{Q}  \tr I^\alpha_{-n} \bm{\texttt{s}} \Lambda W^2
\notag \\
\times \sum_{\beta,m}\int_{\bm{x}^\prime}
\tr I^\beta_m \bm{\texttt{s}} W   \tr \bm{\texttt{A}}_{-m}^\beta W 
\Rrangle  \to \frac{\delta_{(1),1;(1),2}\Gamma_t}{\Gamma_t} S_{\rm int}[\underline{Q}]  ,
\end{gather}
where
\begin{align}
 \frac{\delta_{(1),1;(1),2}\Gamma_t}{\Gamma_t} & =  - \frac{24\pi T \gamma}{g D}\sum_{m>0}\int_p \DD_p^t(i\omega_m)\notag \\
 & +
\frac{32\pi T \gamma}{g D} \sum_{m>0} \int_p
\mathcal{D}^{t2}_p(i\omega_m) .
\label{eq:Gammat:11:12}
\end{align}
Last contribution to  $\Gamma_t$ is provided by the following combination
\begin{gather}
 \Llangle \delta S_{\rm int}^{(2),2}+\frac{1}{2} \bigl(\delta S_{\rm int}^{(1),2}\bigr)^2 \Rrangle \!=\! 
 - \frac{\pi T\Gamma_t}{4}\!\!\sum_{\alpha,\beta;nm} \! \sum_{\textsf{j},\textsf{k}=1}^3
\int\limits_{\bm{x},\bm{x}^\prime}\!\!
\Llangle \Bigl ( \delta_{nm}  \notag \\
\times \delta^{\alpha\beta}\delta_{\textsf{j}\textsf{k}}\delta(\bm{x}-\bm{x}^\prime) - 2\pi T \Gamma_t \tr I^\alpha_n \textsf{s}_\textsf{j} W \tr I^\beta_{-m} \textsf{s}_{\textsf{k}} W \Bigr )
\notag \\
\times 
\tr \texttt{A}_{-n,\textsf{j}}^\alpha W  \tr \texttt{A}_{m,\textsf{k}}^\beta W  \Rrangle \to
 \frac{\delta_{(2),2}\Gamma_t}{\Gamma_t} S_{\rm int}[\underline{Q}] ,
 \end{gather}
 where
 \begin{align}
\frac{\delta_{(2),2}\Gamma_t}{\Gamma_t}=  \frac{2}{g}\int_p \mathcal{D}_p(0) -
\frac{16\pi T \gamma}{g D} \sum_{m>0} \int_p\mathcal{D}^{t2}_p(i\omega_m) .
\label{eq:Gammat:22}
\end{align}
Combining all the above contributions  (cf. Eqs. \eqref{eq:Gammat:21}, \eqref{eq:Gammat:11:11}, \eqref{eq:Gammat:11:12}, and \eqref{eq:Gammat:22}) together, we find
\begin{align}
\frac{\delta \Gamma_t}{\Gamma_t} = & -  \frac{2(\textsf{v}-1)}{g}\int_p \mathcal{D}_p(0) 
- \frac{8\pi T \gamma}{g D}\sum_{n>0}\int_p \mathcal{D}^2_p(2i\varepsilon_n) 
\notag \\
& +
\frac{16\pi T \gamma}{g D} \sum_{m>0} \int_p\mathcal{D}_p^2(i\omega_m) 
\to (1-3\gamma) \frac{t h^\epsilon}{\epsilon}  .
\label{eq:renorm:Gammat}
\end{align}
Here the final expression of the last line  is obtained by setting $T=0$ and using $h$ as an infrared regulator together with dimensional regularization in $d=2+\epsilon$ dimension. Parameter $t$ controlling disorder is defined as 
\begin{equation}
t = \frac{2^{2-d}\Gamma(2-d/2)}{g\pi^{d/2}} \qquad \stackrel{d=2}{\longrightarrow} \qquad t=\frac{1}{\pi g}\ .
\end{equation}

\subsection{Background field renormalization of $Z_\omega$}

Now we compute the one-loop renormalization of $Z_\omega$. We start from the following contribution
\begin{gather}
\langle \delta S_{\rm 0}^{(\varepsilon)}\rangle = -\frac{Z_\omega}{2}  \int\limits_{\bm{x}} \Tr [\TT \hat \varepsilon,\texttt{T}^{-1}] \Lambda \langle W^2\rangle \to
\delta_\varepsilon Z_\omega \Tr  \hat \varepsilon\underline{Q} ,
\end{gather}
where (cf. Eq. \eqref{eq:Gammat:21})
\begin{gather}
\frac{\delta_\varepsilon Z_\omega}{Z_\omega} = -\frac{\textsf{v}}{g}
 \int_p \mathcal{D}_p(0)  + \frac{12\pi T\gamma}{gD}\sum_{n>0} \int_p \DD^t_p(i\omega_n)  .
 \label{eq:deltaZ:eps}
\end{gather}
The second contribution comes from
\begin{gather}
\frac{1}{2}\Llangle \bigl(\delta S_{\rm int}^{(1);2}\bigr )^2\Rrangle = \frac{\left (\pi T\Gamma_t\right)^2}{8}  \Llangle \Bigl ( \sum_{\alpha, n}
\int\limits_{\bm{x}} \tr I_n^\alpha \bm{\textsf{s}}W \notag \\
\times \tr \bm{\texttt{A}}_{-n}^\alpha W \Bigr )^2 \Rrangle
\to \frac{2 \pi T\Gamma_t^2}{g} \int\limits_{\bm{x}\bm{x}^\prime} 
\sum_{\alpha, \omega_n>\Omega} \omega_n \mathcal{D}_{\bm{x}-\bm{x}^\prime}^t(i\omega_n) \notag \\
\times
\langle \tr \bm{\texttt{A}}_{n}^\alpha(\bm{x}) W(\bm{x})
\tr \bm{\texttt{A}}_{-n}^\alpha(\bm{x}^\prime) W(\bm{x}^\prime)  \rangle . 
\end{gather}
Here we singled out the contribution from large values of $n$. The part with $\omega_n<\Omega$ does not contribute to the renormalization of $Z_\omega$. 
We note that due ``largeness'' of $\omega_n$, either $\TT$ or $\texttt{T}^{-1}$ in the expression for $\bm{\texttt{A}}_{n}^\alpha$ is the identity matrix. Then we find
\begin{gather}
\frac{1}{2}\Llangle \! \bigl(\!\delta S_{\rm int}^{(1);2}\!\bigr )^2\!\Rrangle \!\to\! 
\frac{12 \pi T\gamma \Gamma_t}{g D} \! \int\limits_{\bm{x}\bm{x}^\prime} \!
\sum_{\omega_n>0} \omega_n \mathcal{D}_{\bm{x}-\bm{x}^\prime}^t(i\omega_n)  
 \! \! \! \sum_{\alpha,\beta;k,m}\notag \\
\times \mathcal{D}_{\bm{x}-\bm{x}^\prime}(i\omega_{n+|m|}-i \omega_k \sgn \omega_m)
\tr \TT_{m k}^{\beta\alpha}(\bm{x})[\texttt{T}^{-1}(\bm{x}^\prime))]_{km}^{\alpha\beta} .
\label{eq:Sint:2}
\end{gather}
Expanding the propagator to the first power in ``small'' frequencies $\omega_k$ and $\omega_m$, we obtain 
\begin{gather}
\frac{1}{2}\Llangle \bigl(\delta S_{\rm int}^{(1);2}\bigr )^2\Rrangle \to 
\delta_{(1);2;(1);2)} Z_\omega \Tr  \hat \varepsilon\underline{Q},
\end{gather}
where
\begin{align}
\frac{\delta_{(1);2;(1);2)} Z_\omega}{Z_\omega}=
\frac{12\pi T \gamma}{g D}
\sum_{n>0} \int_p \Bigl [  \mathcal{D}^2_{p}(i\omega_{n})-\DD_{p}^t(i\omega_n)\Bigr ]  .
\label{eq:deltaZ:12}
\end{align}
Combining together both contributions, cf. Eqs. \eqref{eq:deltaZ:eps} and \eqref{eq:deltaZ:12}, we find
\begin{gather}
\frac{\delta Z_\omega}{Z_\omega}= -
\frac{\textsf{v}}{g} \int_p \mathcal{D}_p(0)  + \frac{12\pi T\gamma}{g D}\sum_{n>0} \int_p \mathcal{D}^2_p(i\omega_n) 
\notag \\
\to (1-3\gamma)\frac{t h^\epsilon}{\epsilon} .
\label{eq:ren:Z}
\end{gather}
We note that inspection of Eqs. \eqref{eq:renorm:Gammat} and  \eqref{eq:ren:Z} demonstrates that $\delta Z_\omega/Z_\omega \equiv \delta \Gamma_t/\Gamma_t$ within the one-loop approximation (the lowest order in $t$). It implies that the dimensionless interaction parameter is not renormalized,  
\begin{equation}
\delta\gamma\equiv 0 .
\label{eq:ren:gamma}
\end{equation}

\subsection{Background field renormalization of $g$}

We start from the following comment. The matrix vector field $\bm{\texttt{A}}$ is related with the matrix $\underline{Q}$ as: $\Tr [\bm{\texttt{A}},\Lambda]^2=\Tr (\nabla \underline{Q})^2$. Therefore, components of $\bm{\texttt{A}}$ anticommuting with $\Lambda$ can only contribute to renormalized effective action. Thus, for a sake of simplicity we can assume that 
 $\bm{\texttt{A}}\Lambda=-\Lambda \bm{\texttt{A}}$. 
 
There are two contributions to renormalization of $g$. At first, we have 
\begin{gather}
\langle \delta S_{\rm 0}^{(2),1} \rangle = \frac{g}{8}\int\limits_{\bm{x}} \Tr \bm{\texttt{A}} \Lambda  \bm{\texttt{A}} \Lambda \langle  W^2 \rangle
\to -\frac{\delta_{(2),1} g}{16} \int\limits_{\bm{x}} \Tr (\nabla \underline{Q})^2  ,
\end{gather}
where (cf. Eq. \eqref{eq:Gammat:21})
\begin{gather}
\frac{\delta_{(2),1} g}{g} = -\frac{\textsf{v}}{g}
 \int_p \mathcal{D}_p(0)  + \frac{12\pi T\gamma}{gD}\sum_{n>0} \int_p \DD^t_p(i\omega_n)  .
 \label{eq:deltag:21}
\end{gather}

The other contribution comes from $\frac{1}{2}\llangle  \bigl(\delta S_{\rm int}^{(1);2}\bigr )^2\rrangle$.  Using Eq. \eqref{eq:Sint:2}, we write
\begin{gather}
\frac{1}{2}\Llangle \! \bigl(\delta S_{\rm int}^{(1);2}\bigr )^2\!\Rrangle
\to \frac{12 \pi T\gamma \Gamma_t}{g D} \int\limits_{\bm{x}\bm{x}^\prime} 
\sum_{\omega_n>0} \omega_n \DD_{x-x^\prime}^t(i\omega_n) \notag \\
\times \tr \TT(\bm{x}) \texttt{T}^{-1}(\bm{x}^\prime) .
\end{gather}
Here we neglect ``small'' frequencies in comparison with ``large'' frequency $\omega_n$.
Next expanding $\TT(\bm{x})$ and $\texttt{T}^{-1}(\bm{x}^\prime)$ near the point $(\bm{x}+\bm{x}^\prime)/2$ to the second order $\bm{x}-\bm{x}^\prime$ 
and using the identity
\begin{gather}
\int\limits_{\bm{x}} \! \bm{x}^2 \DD_{\bm{x}}^t(i\omega_n) =4\! \int\limits_p \! \mathcal{D}_{p}^t(i\omega_n)  \mathcal{D}^2_{p}(i\omega_n)\Bigl [1 -
\frac{4 p^2}{d} \mathcal{D}_{p}(i\omega_n)\Bigr ] ,
\end{gather}
we find
\begin{gather}
\frac{1}{2}\Llangle \! \bigl(\delta S_{\rm int}^{(1);2}\bigr )^2\!\Rrangle
\to -\frac{\delta_{(1),2;(1),2} g}{16} \int_{\bm{x}} \Tr (\nabla \underline{Q})^2 ,
\end{gather}
where
\begin{gather}
\frac{\delta_{(1),2;(1),2} g}{g} = 
\frac{12 \pi T\gamma}{g D} \int_p  \sum_{n>0} \Bigl [\mathcal{D}^2_{p}(i\omega_n) - \DD_{p}^t(i\omega_n)\Bigr ]\notag \\
\times\Bigl [ 1 - \frac{4 p^2}{d} \mathcal{D}_{p}(i\omega_n)\Bigr ] .
\label{eq:Sint:2:g}
\end{gather}
We note that the above contribution contains the full derivative 
\begin{equation}
\int_p \mathcal{D}^2_{p}(i\omega_n) \Bigl [ 1 - \frac{4 p^2}{d} \mathcal{D}_{p}(i\omega_n)\Bigr ]
= -\frac{1}{4d} \int_p \partial_{p_\mu} \partial_{p_\mu}  \mathcal{D}^2_{p}(i\omega_n).
\end{equation}
This term being full derivative does not contribute to renormalization and can be safely neglected.

Combing both contributions to $g$, cf. Eqs. \eqref{eq:deltag:21} and \eqref{eq:Sint:2:g}, together, we obtain
\begin{gather}
\frac{\delta g}{g} = - \frac{\textsf{v}}{g} \int\limits_p \mathcal{D}_p(0) + \frac{48\pi T\gamma}{d g D} 
\sum_{m>0} \int\limits_p p^2 \mathcal{D}^2_p(i\omega_m) \mathcal{D}^t_p(i\omega_m)
\notag \\
 \to \Bigl [ 1 +6 f(\gamma)\Bigr ]\frac{t h^\epsilon}{\epsilon} ,\quad f(\gamma)=1-\frac{1+\gamma}{\gamma}\ln(1+\gamma) .
 \label{eq:deltag:1loop}
\end{gather}

\subsection{Background field renormalization of $h^2$}

Finally, we discuss the background field renormalization of the $S_{\rm h}$ regulator. This is intimately related with the renormalization of the $Q$ matrix itself, so-called $Z$-factor. As we shall discuss below, the latter is also related with the renormalization of the {\DOS}. There is a single contribution to renormalization of $h^2$:
\begin{gather}
\langle \delta S_{\rm 0}^{(h)} \rangle  \!=\! -\frac{g h^2}{16}\!  \int\limits_{\bm{x}} \! \Tr [\TT \Lambda,\texttt{T}^{-1}] \Lambda \langle W^2\rangle 
\! \to \! \frac{g h^2 \delta Z}{8}\!  \int\limits_{\bm{x}}\! \Tr \Lambda \underline{Q} ,
\end{gather}
where (cf. Eq. \eqref{eq:Gammat:21})
\begin{gather}
\delta Z = -\frac{\textsf{v}}{g}
 \int_p \mathcal{D}_p(0)  + \frac{12\pi T\gamma}{gD}\sum_{n>0} \int_p \DD^t_p(i\omega_n)  
 \notag \\
 \to \Bigl [ 1 -3 \ln(1+\gamma)\Bigr ]\frac{t h^\epsilon}{\epsilon}.
 \label{eq:BGF:Zren}
\end{gather}
Introducing the renormalization of $h^2$ according to $\delta (g h^2) = g h^2 \delta Z$ we find
\begin{equation}
\frac{\delta h^2}{h^2} = \delta Z - \frac{\delta g}{g}=
- 3\Bigl [\ln(1+\gamma) +2f(\gamma)\Bigr ] \frac{t h^\epsilon}{\epsilon} .
\end{equation} 
We note that there is no renormalization of $h^2$ in the absence of interaction, $\gamma=0$.

\subsection{One-loop renormalization}

We introduce the renormalized infrared scale $h^\prime$ and renormalized conductance $g^\prime$ as 
\begin{gather}
h^{\prime 2} = \frac{g h^2 Z}{g^\prime} =
h^2 \Bigl [1 -\frac{b t h^{\epsilon}}{\epsilon}\Bigr ] , \qquad 
g^\prime = g \Bigl [1 + \frac{a_1 t h^{\epsilon}}{\epsilon} \Bigr ] ,\notag \\
a_1 = 1 +6 f(\gamma), \qquad b = 3 \ln(1+\gamma)+6f(\gamma) .
\label{eq:hprime:ren}
\end{gather}
Also we introduce renormalized $Z_\omega$ and $\Gamma_t$:
\begin{gather}
\frac{Z_\omega^\prime}{Z_\omega}=\frac{\Gamma_t^\prime}{\Gamma_t}=  1 + (1-3\gamma)\frac{t h^{\epsilon}}{\epsilon} .
\label{eq:ren:1loop:ZG}
\end{gather}
Then applying the minimal subtraction scheme (see e.g. the book \cite{Amit-book} for details), we can formulate the one-loop renormalization group equations as
\begin{subequations}
\begin{align}
\frac{d t}{d\ell} & = -\epsilon t + \bigl [{\textsf{v}}/{2}+6f(\gamma)\bigr ] t^2 + O(t^3),\\
\frac{d \gamma}{d\ell } &  =  0+ O(t^2) , \\
\frac{d \ln Z_\omega}{d\ell } & = - ({\textsf{v}}/{2}-3\gamma)t  + O(t^2) , \\
\frac{d\ln Z}{d\ell} &  = -\bigl [{\textsf{v}}/{2}- 3 \ln(1+\gamma)\bigr ]t+ O(t^2) .
\end{align}
\label{eq:RG:one-loop}
\end{subequations}
where $\ell= \ln 1/h^\prime$ plays the role of the logarithm of the infrared lengthscale. 
At $T=0$ the later is just a system size. At finite temperature the infrared scale is set by the temperature length $\sim\sqrt{D/T}$.  We remind that $\textsf{v}=2$.

We mention that Eqs. \eqref{eq:RG:one-loop} coincide with renormalization group equations obtained in Refs. \cite{Jeng2001a,Jeng2001,DellAnna2006,Liao2017}.

\section{Linear response\label{Sec:LinearR}}

The background field method allowed us to derive renormalization of the {\NLSM} action in the one-loop approximation (lowest order in disorder). However, this method is not convenient for calculation of the renormalization beyond the one-loop approximation. Such calculation is of particular importance since the dimensionless interaction strength $\gamma$ is not renormalized within the lowest order in $t$ (see discussion in Sec. \ref{Sec:Final}). Therefore, in this section we present an alternative approach to obtain renormalization of parameters of the {\NLSM} action.

\subsection{Dynamical spin susceptibility}

The Finkel'stein {\NLSM} \eqref{eq:NLSM} possesses spin rotational symmetry. Therefore, it is natural to characterize the linear response in class C by the dynamical spin susceptibility. Let us define it as follows ($\omega_n>0$),
\begin{gather}
\chi_{\textsf{j}\textsf{k}}(q,i\omega_n)=Z_s \delta_{\textsf{j}\textsf{k}} - \frac{\pi}{2}T Z_s^2 \langle \Tr I^\alpha_n \textsf{s}_{\textsf{j}} Q_{\bm{q}} \Tr I^\alpha_{-n} \textsf{s}_{\textsf{k}} Q_{\bm{-q}} \rangle .
\label{eq:DynSus:1}
\end{gather}
Here $Q_{\bm{q}}$ denotes the Fourier transform of $Q(\bm{r})$ in the momentum space.
We note that $\alpha$ is a fixed replica index.
The first term on the right hand side of Eq. \eqref{eq:DynSus:1} describes the static part whereas the correlation function takes into account dynamic part. On a tree level we can substitute $W$ for $Q$ in Eq. \eqref{eq:DynSus:1}. Then we find 
\begin{gather}
\chi_{\textsf{j}\textsf{k}}= \chi_s \delta_{\textsf{j}\textsf{k}}, \qquad \chi_s(q,i\omega_n)= 
Z_s - \frac{Z_s^2\omega_n/(Z_\omega)}{Dq^2+ \omega_n (1+\gamma)} .
\label{eq:DynSus:2}
\end{gather}
The spin rotational symmetry (the conservation of the spin) results in the well-known Ward identity, $\chi_s(q=0,i\omega_n\to 0)=0$. The later is consistent with Eq. \eqref{eq:DynSus:2} provided
\begin{equation}
Z_s = Z_\omega+\Gamma_t=Z_\omega(1+\gamma)  .
\label{eq:Zs}
\end{equation}
Then we obtain \footnote{We note the factor $2/\pi$ difference in definition of the dynamical spin susceptibility with Ref. \cite{Liao2017}.}
\begin{equation}
\chi_s(q,i\omega_n)= 
Z_s \frac{Dq^2}{Dq^2+ \omega_n (1+\gamma)} .
\label{eq:DynSus:3}
\end{equation}
We note that we omit the regulator term in the {\NLSM} action ($h=0$) in order to have a standard diffusive pole. The diffusive form of the dynamical spin susceptibility allows one to extract renormalization of the combination $Z_\omega+\Gamma_t$ and the spin conductance,
\begin{subequations}
\begin{align}
Z_\omega^\prime+\Gamma_t^\prime = Z_s^\prime = \lim\limits_{q\to 0} \lim\limits_{\omega_n/q^2\to 0}\chi_s(q,i\omega_n), \\ 
g^\prime = \lim\limits_{\omega_n\to 0} \lim\limits_{q^2/\omega_n\to 0}\frac{16 \omega_n}{q^2} \chi_s(q,i\omega_n) .
\label{eq:def:g:Zs}
\end{align}
\end{subequations}
We note that `prime' sign indicates that these quantities are obtained after calculation of quantum corrections. 

The one-loop correction to the dynamical spin susceptibility is given as 
\begin{align}
\delta \chi_s  & =  - \frac{1}{2}\pi T Z_s^2 \Biggl \{  \frac{1}{4}\langle \tr I^\alpha_n \sz\Lambda W^2_{\bm{q}} \tr I^\alpha_{-n} \sz \Lambda W^2_{\bm{-q}} \rangle \notag \\ 
 + &  \Llangle \tr I^\alpha_n \sz W_{\bm{q}} \tr I^\alpha_{-n} \sz W_{\bm{-q}}
\begin{bmatrix} 
S_{\rm 0}^{(4)}+ S_{\rm int}^{(4)} \\
+\frac{1}{2}(S_{\rm int}^{(3)})^2
\end{bmatrix} \Rrangle\Biggr \} .
\label{eq:DynSpin:1loop}
\end{align}
Here we introduce the following non-Gaussian terms stemming from the expansion of $Q$ matrix in powers of $W$ of the {\NLSM} action:
\begin{gather}
S_{\rm 0}^{(4)}  = \frac{g}{64}\prod_{i=1}^4\int\limits_{q_i} \sum_{\alpha_i,n_i}
\delta\left (\sum_i\bm{q}_i\right)
\Bigl ( 
\bm{q}_{12}\bm{q}_{34}+
\bm{q}_{14}\bm{q}_{23}-2h^2
\notag \\
-  \frac{\omega_{n_{12}+n_{34}}}{D}
\Bigr ) \tr \Bigl [(w_{\bm{q_1}})_{n_1n_2}^{\alpha_1\alpha_2}
(w_{\bm{q_2}}^\dag)_{n_2n_3}^{\alpha_2\alpha_3}
(w_{\bm{q_3}})_{n_3n_4}^{\alpha_3\alpha_4}
(w_{\bm{q_4}}^\dag)_{n_4n_1}^{\alpha_4\alpha_1}
\Bigr ]
\end{gather}
(here we use a shorthand notation $\bm{q}_{12}\equiv \bm{q}_1+\bm{q}_2$) and
\begin{subequations}
\begin{align}
S^{(3)}_{\rm int}  & = \frac{\pi T \Gamma_t}{4}  \sum_{\alpha,n} \int\limits_{\bm{x}} \Tr I^{\alpha}_{n} \bm{\textsf{s}} W
\Tr I^{\alpha}_{-n} \bm{\textsf{s}}\Lambda W^2 ,\\
S^{(4)}_{\rm int} & = -\frac{\pi T\Gamma_t }{16} \sum_{\alpha,n} 
\int\limits_{\bm{x}} \Tr I^{\alpha}_{n} \bm{\textsf{s}} \Lambda W^2
\Tr I^{\alpha}_{-n} \bm{\textsf{s}}\Lambda W^2 .
\end{align}
\end{subequations}
We note that the first line in Eq. \eqref{eq:DynSpin:1loop} contributes to the renormalization of the static part, i.e. the quantity $Z_s$, whereas the second line contains information on renormalization of the spin conductance. 

Evaluating the average in the first line of Eq. \eqref{eq:DynSpin:1loop} (we can set $n$ and $q$ to zero from the very beginning), we find 
\begin{gather}
\delta Z_s\!=\!
\frac{32 \pi T Z_s^2}{g^2}
\!\!\int\limits_p \!\sum_{m>0} \Bigl [
\mathcal{D}_p^2(i2\varepsilon_m)\! +\! 2 \mathcal{D}_p^{t2}(i\omega_m)\!-\! 2\mathcal{D}_p^2(i\omega_m)
\Bigr ] \notag 
\\
\to Z_s(1-3\gamma)\frac{t h^\epsilon}{\epsilon}.
\label{eq:ren:1loop:Zs}
\end{gather}
We note that the result \eqref{eq:ren:1loop:Zs} is fully consistent with Eq. \eqref{eq:ren:1loop:ZG}. Therefore, the renormalization by means of the background field method and via linear response do match each other as expected.

The computation of the correction to the spin conductance is a bit more involved. 
We start from the term with $S_{\rm 0}^{(4)}$ in the second line of Eq. \eqref{eq:DynSpin:1loop}. At first making the partial averaging, we obtain \begin{gather}
S_{\rm 0}^{(4)}  \to \int\limits_{pq}\sum_{\alpha_3\alpha_4;n_3n4} 
\Biggl \{
- \frac{\textsf{v}}{16} \Bigl [
\Bigl ( 
 \mathcal{D}_p^{-1}(i2\varepsilon_{n_3}) +\mathcal{D}^{-1}_q(i\omega_{n_{34}}) \Bigr )
 \notag \\
 \times \mathcal{D}_p(i2\varepsilon_{n_3}) 
+ 
\Bigl (
 \mathcal{D}_p^{-1}(i2|\varepsilon_{n_4}|) +\mathcal{D}^{-1}_q(i\omega_{n_{34}}) \Bigr )\mathcal{D}_p(i2|\varepsilon_{n_4}|) \Bigr ]
\notag \\
+ \frac{3 \pi T \Gamma_t}{g} \Bigl[ \sum_{\omega_n>\varepsilon_{n_3}}+\sum_{\omega_n>|\varepsilon_{n_4}|}\Bigr ] \Bigl [
\mathcal{D}^{-1}_p(i\omega_{n})+
\mathcal{D}^{-1}_q(i\omega_{n_{34}}) \Bigr ] \notag \\
\times
\DD^t_p(i\omega_n)
\Biggr \}
\tr [w_{n_3n_4}^{\alpha_3\alpha_4}(\bm{q})
{w}_{n_4n_3}^{\dag\alpha_4\alpha_3}(-\bm{q}) ] .
\label{eq:S4sigma:part:int}
 \end{gather}
Then we find (in the limit $q^2/\omega_n \to 0$)
\begin{gather}
\frac{1}{2}\pi T Z_s^2 \Llangle\! \tr I^\alpha_n \sz W_{\bm{q}} \tr I^\alpha_{-n} \sz W_{\bm{-q}}
S_{\rm 0}^{(4)}\!\Rrangle\!\to\!
\frac{\textsf{v}q^2}{16 \omega_n}\! \int\limits_p\! \mathcal{D}_p(0) \notag 
 \\
 - \frac{q^2}{16 \omega_n}\frac{12\pi T\gamma}{D}\sum_{m>0} \int\limits_p \DD^t_p(i\omega_m).
 \label{eq:dlg:1}
 \end{gather}
Next we consider the terms in Eq. \eqref{eq:DynSpin:1loop} due to $S_{\rm int}$. After tedious but straightforward calculation we obtain
\begin{gather}
\frac{1}{2}\pi T Z_s^2 \Llangle \tr I^\alpha_n \sz W_{\bm{q}} \tr I^\alpha_{-n} \sz W_{\bm{-q}}
\Bigl[S_{\rm int}^{(4)} 
+\frac{1}{2}(S_{\rm int}^{(3)})^2\Bigr ]
\Rrangle 
\notag \\
\to \frac{q^2}{16 \omega_n} \frac{12\pi T\gamma}{D} \int_p \sum_{m>0}\DD^t_p(i\omega_m)
\Bigl [1 - \frac{4p^2}{d}\mathcal{D}_p(i\omega_m)\Bigr ].
\label{eq:dlg:2}
\end{gather}
Combining both contributions, Eqs. \eqref{eq:dlg:1} and \eqref{eq:dlg:2}, we reproduce the result \eqref{eq:deltag:1loop}.

\subsection{Spin conductance}

Although, as we demonstrated above, dynamical spin susceptibility at finite momentum can be used to extract renormalization of the spin conductance, taking the double limit, cf. Eq. \eqref{eq:def:g:Zs},  is not convenient. In standard symmetry classes one can directly  express conductance in terms of correlation functions of the $Q$-field. 
Such Kubo formula involves the matrix current $Q\nabla Q$. In the present case the most general form of the current can be written as $\Tr I_n^\alpha \textsf{s}_\textsf{j} Q\nabla Q$. Applying the symmetry \eqref{eq:symm:C}, we find
\begin{equation}
\Tr I_n^\alpha \textsf{s}_\textsf{j} Q\nabla Q = \textsf{v}_\textsf{j} 
\Tr I_n^\alpha \textsf{s}_\textsf{j} Q\nabla Q .
\end{equation}
Thus the current operator corresponding to the charge current vanishes identically, $\Tr I_n^\alpha Q\nabla Q\equiv 0$. However, the spin current $\Tr I_n^\alpha \textsf{s}_\textsf{j} Q\nabla Q$ is nontrivial. Using SU(2) spin symmetry we write the Kubo formula for the spin conductivity as (similar expression has been proposed in Ref. \cite{DellAnna})
\begin{gather}
\sigma_s(i\omega_n) = - \frac{g}{8 n} \bigl \langle \Tr [I_n^\alpha \sz,Q][I_{-n}^\alpha \sz,Q]\bigr \rangle
+\frac{g^2}{16 d n} \int\limits_{\bm{x}^\prime} \notag \\
\times  \!\Llangle\!
\Tr I_n^\alpha \sz Q(\bm{x}) \nabla Q(\bm{x}) 
\Tr I_{-n}^\alpha \sz Q(\bm{x}^\prime) \nabla Q(\bm{x}^\prime)\!
\Rrangle  .
\label{eq:sigmaS:def}
\end{gather} 
Here the first term in the right hand side of Eq. \eqref{eq:sigmaS:def} describes the so-called diamagnetic contribution whereas the second corresponds to  the current-current correlation. Using the relation $\Tr [I_n^\alpha,\Lambda][I_{-n}^\alpha,\Lambda]=-8n$, we find the bare value of the spin conductivity as $\sigma_s(i\omega_n)=g$. Therefore, we expect that $\sigma_s(i\omega_n)$ in the static limit $\omega_n\to 0$ gives the renormalized conductance.

We start from diamagnetic-like contribution. To the lowest nontrivial order in $W$ it reads 
\begin{gather}
\frac{g}{8 n} \bigl \langle \Tr [I_n^\alpha \sz,Q][I_{-n}^\alpha \sz,Q]\bigr \rangle
\simeq \frac{g}{8 n} \Bigl \langle \Tr \Bigl [2  I_n^\alpha \sz W I_{-n}^\alpha \sz W
\notag \\ 
- (I_n^\alpha \Lambda I_{-n}^\alpha 
+I_{-n}^\alpha \Lambda I_{n}^\alpha)\Lambda W^2  \Bigr ] \Bigr \rangle
 =\frac{\textsf{v}}{n} \int\limits_q \sum_{\omega_n>\varepsilon_m>0}
\mathcal{D}_q(2i\varepsilon_m)
\notag \\
{-} \frac{1}{n} \!\int\limits_q\! \sum_{\varepsilon_m>0}
\Bigl [\textsf{v} \mathcal{D}_q(2i\varepsilon_m)
{+} 4 \mathcal{D}_q(2i\varepsilon_m{+}i\omega_n)  
{+} \textsf{v}  \mathcal{D}_q(2i\varepsilon_{m+n}) 
\Bigr ]
\notag \\ 
+ \frac{\gamma}{n D} \int\limits_q \sum_{\Omega_m>0}
\Bigl [10 \Omega_m \DD^t_q(i\Omega_{m+n})+6\Omega_{m-n}\DD^t_q(i\Omega_m)
\Bigr ]  
 .
\label{eq:g:ren:Kubo:1}
\end{gather}
We note that  the interaction part of the propagator contributes to the diamagnetic-type term contrary to the conductance in the standard Wigner-Dyson classes. 
The other contribution becomes
\begin{gather}
\frac{g^2}{16 d n} \int\limits_{\bm{x}^\prime} \Llangle
\tr I_n^\alpha W(\bm{x}) \nabla W(\bm{x}) 
\tr I_{-n}^\alpha W(\bm{x^\prime}) \nabla W(\bm{x^\prime})
\Rrangle \notag \\
= 
{-}\frac{8}{n d} \!\int\limits_q\! q^2 \!\sum_{\varepsilon_m>0}
\Bigl [\mathcal{D}_q(2i\varepsilon_m){+}\mathcal{D}_q(2i \varepsilon_{m+n})\Bigr ]
\mathcal{D}_q(2i\varepsilon_m{+}i\omega_n)
\notag \\
+ \frac{8 \gamma}{n d D} \int\limits_q \sum_{\Omega_m>0}\Omega_m
\Bigl [\DD_q^t(i\Omega_{m+n})\mathcal{D}_q(i\Omega_{m+2n})
\notag\\
+3 \DD_q^t(i\Omega_{m})\mathcal{D}_q(i\Omega_{m+n})
+4 \mathcal{D}_q^t(i\Omega_{m})\DD_q^t(i\Omega_{m+n})
\Bigr] .
\label{eq:g:ren:Kubo:2}
\end{gather}
We note that each of the contributions \eqref{eq:g:ren:Kubo:1} and \eqref{eq:g:ren:Kubo:2} diverges in the limit $\omega_n\to 0$ but their sum 
is finite. Expanding in $\omega_n$, we find the static spin conductivity
\begin{gather}
\sigma_s^\prime = g + \frac{8}{n d} \!\int\limits_q\! \partial_{q_\mu}\Bigl \{q_\mu \Bigl [\sum_{\varepsilon>0}
 \mathcal{D}_q(2i\varepsilon)
- \frac{2 \gamma}{D} \sum_{\Omega>0}\Omega
\DD^t_q(i\Omega)\Bigr ]\Bigr\}
\notag \\
+ \frac{32\pi T \gamma}{d D} \int\limits_q \sum_{\Omega>0}
\partial_{q_\mu}\Bigl [q_\mu \DD^t_q(i\Omega)\Bigr ]
\notag \\
- \frac{\textsf{v}}{g} \int\limits_q \mathcal{D}_q(0)    
+ \frac{48 \pi T\gamma}{d D}\int\limits_q q^2 \sum_{\Omega>0}\mathcal{D}^2_q(i\Omega)
\mathcal{D}^t_q(i\Omega) .
\end{gather}
Neglecting the terms which are full derivatives in the first and second line, we obtain $\sigma_s^\prime =g +\delta g$ where $\delta g$ is given by Eq. 
\eqref{eq:deltag:1loop}. We note that Kubo formula \eqref{eq:sigmaS:def} can be more convenient than dynamical spin susceptibility in order to study the renormalization of the spin conductance beyond one-loop approximation.

\section{Local derivativeless operators\label{Sec:Operators}}

\subsection{General construction}

In this section we construct the local pure scaling operators without spatial derivatives. These operators  are eigenoperators with respect to the renormalization group, i.e. 
the renormalization group flow preserves their form. We shall follow the approach of Ref. \cite{Repin2016}.

The simplest local derivativeless operator is related with the {\DOS}. It can be written as
\begin{equation}
\mathcal{K}_1(E) = \frac{1}{4} \sum_{p=\pm} \mathcal{P}_1^{\alpha;p}(E) .
\label{eq:K1}
\end{equation}
Here the retarded/advanced correlation function $\mathcal{P}_1^{\alpha;\pm}(E)$ is defined from its Matsubara counterpart
\begin{equation}
P^\alpha_1(i\varepsilon_n) = \tr \langle Q_{nn}^{\alpha\alpha}\rangle 
\label{eq:P1M}
\end{equation}
as a result of a standard analytic continuation, $i\varepsilon_n \to E+ip 0^+$. 
We emphasize that a replica index $\alpha$ and Matsubara energy index $n$ are fixed. Since the operator $\mathcal{K}_1(E)$ corresponds to the disorder-average {\DOS}, it stays invariant  under the action of the renormalization group. We shall demonstrate this statement explicitly below. 

Next we turn to the local operators without derivatives with two $Q$ matrices. Let us introduce 
\begin{equation}
\mathcal{K}_2(E_1,E_2) = \frac{1}{16}\sum_{p_1,p_2=\pm} p_1p_2 \mathcal{P}_2^{\alpha_1\alpha_2;p_1p_2}(E_1,E_2) ,
 \label{eq:K2}
 \end{equation}
where the correlation function $\mathcal{P}_2^{\alpha_1\alpha_2;p_1p_2}(E_1,E_2)$ is related with its Matsubara counterpart
 \begin{gather}
 P_2^{\alpha_1\alpha_2}(i\varepsilon_{n},i\varepsilon_{m}) =\bigl  \langle \tr Q_{nn}^{\alpha_1\alpha_1}(\bm{r}) \tr Q_{mm}^{\alpha_2\alpha_2}(\bm{r}) \bigr \rangle \notag \\
   +\mu_2  \bigl \langle \tr \bigl [Q_{nm}^{\alpha_1\alpha_2}(\bm{r}) Q_{mn}^{\alpha_2\alpha_1}(\bm{r}) \bigr ] \bigr \rangle 
 \label{eq:P2M}
 \end{gather}
by standard analytic continuation to the real frequencies:  $\varepsilon_{n} \to E_1+i p_10^+ $ and $\varepsilon_{m} \to E_2+i p_20^+ $. We note that no summation over $\alpha_1, \alpha_2, n,$ and $m$ is assumed and $\alpha_1\neq\alpha_2$. The latter inequality reflects the fact that we are interested in mesoscopic fluctuations in the presence of interaction. We mention that if one interested in the renormalization of operator \eqref{eq:K2} alone then one can use the following simplified definition
\begin{equation}
K_2 = \frac{1}{16}\lim\limits_{\varepsilon_n,\varepsilon_m\to 0}\sum_{p_1,p_2=\pm} p_1p_2 P_2^{\alpha_1\alpha_2}(i p_1 |\varepsilon_n|,i p_2 |\varepsilon_m|) ,
 \label{eq:K2:simp}
 \end{equation}

The operator $\mathcal{K}_2(E_1,E_2)$ depends on a parameter $\mu_2$. There are particular (integer) values of $\mu_2$ for which $\mathcal{K}_2(E_1,E_2)$ becomes the eigenoperator under the action of renormalization group. 

An operator which involves the number $q$ of matrix fields $Q$ can be constructed in a similar way as above.  We introduce
\begin{gather}
\mathcal{K}_q(E_1,\dots,E_q) = \frac{1}{4^{q}} \sum_{p_1,\dots p_q =\pm} \left ( \prod\limits _{j=1}^q p_j\right )
\notag \\
\times
\mathcal{P}_q^{\alpha_1,\dots,\alpha_q;p_1,\dots,p_q}(E_1,\dots,E_q)  ,
 \label{eq:Kq}
 \end{gather}
where $\mathcal{P}_q^{\alpha_1,\dots,\alpha_q;p_1,\dots,p_q}(E_1,\dots,E_q)$ is related with the Matsubara correlation function $P_q^{\alpha_1,\dots,\alpha_q}(i\varepsilon_{n_{1}},\dots,i\varepsilon_{n_{q}})$ by the analytic continuation to the real frequencies:  $\varepsilon_{n_j} \to E_j+i p_j 0^+$. The later is given as
\begin{gather}
 P_q^{\alpha_1,\dots,\alpha_q}(i\varepsilon_{n_{1}},\dots,i\varepsilon_{n_{q}}) = 
 \sum\limits_{\{k_1,\dots,k_q\}} \mu_{k_1,\dots, k_s} \bigl \langle  R_{k_1,\dots, k_s} \bigr \rangle ,
 \notag \\
 R_{k_1,\dots, k_s}=
 \prod\limits_{r=k_1}^{k_s}  
   \tr Q_{n_{j_1}n_{j_2}}^{\alpha_{j_1} \alpha_{j_2}} Q_{n_{j_2}n_{j_3}}^{\alpha_{j_2}\alpha_{j_3}}\dots Q_{n_{j_r}n_{j_1}}^{\alpha_{j_r}\alpha_{j_1}} .
   \label{eq:PqM}
 \end{gather}
The summation in the right hand side of Eq. \eqref{eq:PqM} is performed over all partitions of the integer number $q$, i.e. over all sets of positive integer numbers $\{k_1,\dots,k_s\}$ which satisfy the following conditions: $k_1+k_2+\dots k_s =q$ and $k_1\geqslant k_2 \geqslant \dots \geqslant k_s>0$. As above all replica indices are different:
$\alpha_j \neq \alpha_k$ if $j\neq k$ for $j,k = 1, \dots, q$. One coefficient among the set $\{\mu_{k_1,\dots, k_s}\}$ can be chosen arbitrary. We stick to the normalization: 
\begin{equation}
\mu_{1,1,\dots,1} = 1 .
\end{equation} 

As we shall see below, for a given $q$ the number of eigenoperators coincide with the number of partitions $(k_1,\dots,k_s)$. Therefore, it will be convenient  to denote the eigenoperators by the partitions $(k_1,\dots,k_s)$ of the integer number $q$ (see details in Ref.~\cite{Karcher2021}).

In the absence of interaction, $\Gamma_t=0$, the {\NLSM} action reduces to Eq. \eqref{eq:S:int}. Since the Matsubara indices of the $Q$ matrix are not mixed without interaction (the energy of diffusive modes conserves), one can project $Q$ matrix to the $2\times 2$ subspace of a given positive and a given negative Matsubara frequencies. Then the group $G$ reduces to $\tilde{G}={\rm Sp}(4 N_r)$ and the effective action becomes $K$-invariant, i.e. invariant under rotations $Q \to \texttt{T}^{-1} Q \texttt{T}$ with $\texttt{T} \in \tilde{K} = {\rm U}(2N_r)$. Then operators $\mathcal{K}_q$ can be averaged over $K$ rotations and resulting $K$-invariant operators can be classified with respect to the irreducible representations of $\tilde{G}$. Each irreducible representation contains single $K$-invariant pure scaling operator \cite{Wegner1986,Gruzberg2013,Karcher2021}. We note that in the noninteracting case one can use also the highest weight vectors approach or the Iwasawa decomposition in order to construct eigenoperators with respect to the renormalization group transformation \cite{Gruzberg2013,Karcher2021}. 

The presence of interaction in the {\NLSM} action introduces several complications for the approach of construction of pure scaling operators 
developed in Refs. \cite{Gruzberg2013,Karcher2021}. At first, we have to deal with fermionic representation of the {\NLSM}. We note that works \cite{Gruzberg2013,Karcher2021} deal with bosonic realisation of the {\NLSM}. 
As discussed in Refs. \cite{Gruzberg2013,Karcher2021}, to extend their analysis to fermionic {\NLSM} is far from being obvious. Secondly,  it is not clear how to extend the classification of $\mathcal{K}_q$ operators with respect to the irreducible representations of the group $G$. For a given $N_m$ this group is finite but we need to take  the limit $N_m\to \infty$. 
In third, the {\NLSM} action is no more $K$-invariant, that is why we have to work with non-$K$-invariant operators. 
Because of these circumstances we employ two complementary approaches. In order to determine the pure scaling operators (to fix the set of the coefficients $\{\mu_{k_1,\dots,k_s}\}$) we shall employ the background field renormalization. In order to check that the structure of the  eigenoperators is not affected by the interaction we shall perform the two-loop renormalization procedure. The latter allows us to see the effect of interaction on anomalous dimensions of pure scaling operators.

\subsection{Operator with single $Q$ matrix}

We start analysis  from the operator with single $Q$ matrix. At first, we shall perform the background field renormalization of this operator in order to demonstrate that it is the eigenoperator indeed. As we shall see, the presence of interaction affects the renormalization of this operator already at the one-loop approximation.

Employing the background field method, we find
\begin{gather}
P^\alpha_1(i\varepsilon_n) \to  \tr \langle [\texttt{T}^{-1} Q \texttt{T}]_{nn}^{\alpha\alpha}\rangle = \Tr \texttt{T} \texttt{P}_{n}^\alpha \texttt{T}^{-1} \langle Q \rangle  \notag \\
= Z(i\varepsilon_n) \tr \underline{Q}_{nn}^{\alpha\alpha}  ,
\label{eq:P1:1}
\end{gather}
where we introduced the projection operator to the fix replica and Matsubara energy,
\begin{equation}
(\texttt{P}_{n}^\alpha)^{\beta\gamma}_{mk} =\delta^{\alpha\beta}\delta^{\alpha\gamma}\delta_{nk}\delta_{nm} \so . 
\end{equation} 
and  (cf. Eq. \eqref{eq:BGF:Zren})
\begin{align}
Z(i\varepsilon_n) & =  1 -\frac{\textsf{v}}{g} \int_p \mathcal{D}_p(2i|\varepsilon_n|) 
\notag \\
& +  \frac{12\pi T\gamma}{g D} \int_p \sum_{\omega_m>|\varepsilon_n|} 
\DD_p^t(i\omega_m) .
\label{eq:P1:2}
\end{align} 
As we see from Eq. \eqref{eq:P1:1}, the operator $K_1$ is the eigenoperator under the action of the renormalization group. Using Eqs. \eqref{eq:P1:1} and \eqref{eq:P1:2}, we obtain
\begin{gather}
K_1 = Z K_1[\Lambda], \quad  Z = 
1 +\left (\frac{\textsf{v}}{2} - 3 \ln(1+\gamma)\right )  
 \frac{t h^{\epsilon}}{\epsilon} \notag \\
 = 1 +\left (\frac{\textsf{v}}{2} - 3 \ln(1+\gamma)\right )  
 \frac{t h^{\prime \epsilon}}{\epsilon}  .
\end{gather}
Applying the minimal subtraction scheme, we deduce the anomalous dimension of the operator $K_1$,
\begin{equation}
\eta_{\textsf{(1)}} = -\frac{d\ln Z}{d\ell} = \left [1 - 3 \ln(1+\gamma)\right ] t + O(t^2) .
\label{eq:Z:1loop}
\end{equation}
The interaction affects the anomalous dimension of $Z$ in the one-loop approximation. Therefore, we shall not compute its two-loop contribution here. As we shall see below, in order to  perform two-loop renormalization of operators involving $q\geqslant 2$ $Q$ matrices, one-loop result \eqref{eq:Z:1loop} is enough.

\subsection{Operators with two $Q$ matrices}

Now we move on to the eigenoperators  with two $Q$ matrices. At first, in order to find them, we shall perform the background field renormalization. However, as we shall see, the background field renormalization is insensitive to the electron-electron interaction. Therefore, we shall also employ two-loop renormalization of the corresponding eigenoperators.

\subsubsection{Background field renormalization}

We start from the operator with two traces in Eq. \eqref{eq:P2M}, 
\begin{gather}
\tr Q_{nn}^{\alpha\alpha} \tr Q_{mm}^{\beta\beta} \to 
\langle \tr [{\texttt{T}}^{-1} Q {\TT} \TP_{n}^{\alpha}] \tr [{\texttt{T}}^{-1} Q {\TT} \TP_{m}^{\beta}]\rangle
\notag \\
\to \tr \underline{Q}_{nn}^{\alpha\alpha} \tr \underline{Q}_{mm}^{\beta\beta}
-\frac{1}{2}  \tr\underline{Q}_{nn}^{\alpha\alpha}  \tr [ {\texttt{T}}^{-1} \Lambda \langle W^2 \rangle {\TT} \TP_{m}^{\beta}]
\notag \\
+ \langle \tr [W {\TT} \TP_{n}^{\alpha} {\texttt{T}}^{-1}
] \tr [W {\TT} \TP_{m}^{\beta}{\texttt{T}}^{-1} ]\rangle 
\notag \\
-\frac{1}{2}   \tr [ {\texttt{T}}^{-1} \Lambda \langle W^2 \rangle {\TT} \TP_{n}^{\alpha}] \tr \underline{Q}_{mm}^{\beta\beta} \simeq Z^2 \tr \underline{Q}_{nn}^{\alpha\alpha} \tr \underline{Q}_{mm}^{\beta\beta}\notag \\
+ \langle \tr [W {\TT} \TP_{n}^{\alpha} {\texttt{T}}^{-1}
] \tr [W {\TT} \TP_{m}^{\beta}{\texttt{T}}^{-1} ]\rangle .
\end{gather}
Important property of the average over $W$ in the above equation is that both Matsubara indices of matrix ${\TT} \TP_{m}^{\beta}{\texttt{T}}^{-1}$ should be small since the rotation ${\TT}$ represents slow mode. Therefore, the interaction part of the propagator \eqref{eq:prop:full} can be omitted since it does not lead to infrared divergent terms. Then it is straightforward to derive the following identity (for slow matrices $A$ and $B$):
\begin{gather}
\langle \tr A w \tr B {w}^\dag \rangle \simeq 2 Y \tr  \Bigl [\Lambda_- A \Lambda_+ ( B
{-} \bar{B})\Bigr ]  , \quad Y = {-} \frac{t h^\epsilon}{\epsilon} .
\label{eq:prop:rel:1}
 \end{gather}
where $\Lambda_\pm = (1\pm \Lambda)/2$ are projectors onto the subspace of positive and negative Matsubara energies. 
Using Eq. \eqref{eq:prop:rel:1}, we obtain
 \begin{gather}
\langle \tr Q_{nn}^{\alpha\alpha} \tr Q_{mm}^{\beta\beta}\rangle \to
Z^2 \tr \underline{Q}_{nn}^{\alpha\alpha} \tr \underline{Q}_{mm}^{\beta\beta}
- Y \tr \underline{Q}_{nm}^{\alpha\beta} \underline{Q}_{mn}^{\beta\alpha}
\notag \\
+ Y  \tr \underline{Q}_{n,-m}^{\alpha\beta} \underline{Q}_{-mn}^{\beta\alpha} .
\label{eq:O1:ren}
\end{gather}

The background field renormalization of the operator with single trace reads
\begin{gather}
\tr Q_{nm}^{\alpha\beta} Q_{mn}^{\beta\alpha} \to
\langle \tr Q \TT \TP_{n}^{\alpha} \texttt{T}^{-1} Q \TT \TP_{m}^{\beta} \texttt{T}^{-1}\rangle
\to \tr \underline{Q}_{n,m}^{\alpha\beta} \underline{Q}_{mn}^{\beta\alpha}
\notag \\
-\frac{1}{2}  \tr \texttt{T}^{-1} \Lambda \langle W^2\rangle \TT \Bigl [\TP_{n}^{\alpha} \texttt{T}^{-1}\Lambda \TT \TP_{m}^{\beta}
+\TP_{m}^{\beta} \texttt{T}^{-1}\Lambda \TT \TP_{n}^{\alpha}\Bigr] 
  \notag \\
+ \langle \tr W \TT \TP_{n}^{\alpha} \texttt{T}^{-1} W \TT \TP_{m}^{\beta} \texttt{T}^{-1}\rangle \simeq Z^2 \tr \underline{Q}_{n,m}^{\alpha\beta} \underline{Q}_{mn}^{\beta\alpha}
\notag \\
 + \langle \tr W \TT \TP_{n}^{\alpha} \texttt{T}^{-1} W \TT \TP_{m}^{\beta} \texttt{T}^{-1}\rangle .
\end{gather}
Now we shall use the following identity (we assume $A$ and $B$ being slow matrices as above)
\begin{gather}
\langle \tr A w B w^\dag \rangle \!\simeq\!  2 Y \Bigl [ \tr  \Lambda_+ A \tr \Lambda_- B
+ \tr  \Lambda_+ A\Lambda_+\bar{B}\Bigr ] .
\label{eq:prop:rel:2}
\end{gather}
We note that in derivation of Eq. \eqref{eq:prop:rel:2} the following relations were used:
\begin{gather}
\sum_{\textsf{j}} \tr \sk \sj \sk \sj = 8 \delta_{\textsf{k}\textsf{0}}, \quad 
\sum_{\textsf{j}} \textsf{v}_{\textsf{j}} \tr \sk \sj \sk \sj = -4 \textsf{v}_{\textsf{k}} .
\end{gather} 
Using Eq. \eqref{eq:prop:rel:2}, we obtain 
\begin{gather}
\tr Q_{nm}^{\alpha\beta} Q_{mn}^{\beta\alpha} \to
Z^2 \tr \underline{Q}_{n,m}^{\alpha\beta} \underline{Q}_{mn}^{\beta\alpha}
- Y \tr \underline{Q}_{nn}^{\alpha\alpha} \tr \underline{Q}_{mm}^{\beta\beta}
\notag \\
+ Y \tr \underline{Q}_{n,-m}^{\alpha\beta} \underline{Q}_{-mn}^{\beta\alpha} .
\label{eq:O2:ren}
\end{gather}
We emphasize that as follows from Eqs. \eqref{eq:O1:ren} and \eqref{eq:O2:ren}, the background field renormalization  mixes not only operators with one and two traces but also the ones with positive and negative Matsubara frequencies. The later is the consequence of the additional symmetry \eqref{eq:symm:C}. 
Finally, using Eqs. \eqref{eq:O1:ren} and \eqref{eq:O2:ren}, we find
\begin{gather}
K_2[Q] \to Z^2 K_2[\underline{Q}]   - \mu_2 Y  O_{1,1}[\underline{Q}] - (2+\mu_2) Y O_2[\underline{Q}] \notag \\
\simeq
 Z^2 (1-\mu_2 Y) \Bigl [ O_{1,1} + (\mu_2  +(\mu_2^2-\mu_2-2)Y ) O_2 \Bigr ] .
\label{eq:K2:BGFR}
\end{gather}
Here we introduce generalization of the operator $R_{k_1,\dots, k_q}$ in a way similar to transformation from $P_2^{\alpha\beta}(i\varepsilon_n,i\varepsilon_m)$ to $K_2$, cf. Eq.  \eqref{eq:K2:simp}:
\begin{gather}
O_{k_1,\dots, k_q} = 
\frac{1}{4^q} \prod_{j=1}^q \left (\lim\limits_{\varepsilon_{n_j}\to 0}\sum_{p_j=\sgn\varepsilon_{n_j}}  p_j \right )   R_{k_1,\dots,k_q} .
\end{gather}
As follows from Eq. \eqref{eq:K2:BGFR}, in order combination $O_{1,1}+\mu_2 O_2$ serves as the eigenoperator under renormalization group transformation, $\mu_2$ should solve the equation
\begin{gather}
\mu_2^2-\mu_2=2  \quad \Longrightarrow \quad \mu_2= 2,-1 .
\end{gather}  
Therefore, we find two eigenoperators corresponding to $\mu_2=2$ and $\mu_2=-1$. As it follows from Eq. \eqref{eq:K2:BGFR}, the renormalization of these eigenoperators are immune to interaction within one-loop approximation. 

We note that the eigenoperators constructed in Ref. \cite{Karcher2021} differ from ones in this work by the sign of $\mu_2$. This difference in sign is explained by usage of bosonic replica in 
Ref. \cite{Karcher2021} whereas we are employing fermionic replica. Translation from one approach to the other can be done by the sign change of all trace (``$\tr$'') operations.

\subsubsection{One-loop renormalization}

After construction of the eigenoperators with two $Q$ matrices with the help of the background field renormalization method we study how the interaction affects their anomalous dimensions. It will be convenient to consider the irreducible part of the correlation function \eqref{eq:P2M},
\begin{equation}
P_2^{\alpha\beta; {\rm (irr)}}(i\varepsilon_n,i\varepsilon_m) \!=\! \llangle \tr Q_{nn}^{\alpha\alpha} \tr Q_{mm}^{\beta\beta} \rrangle + \mu_2 \langle \tr Q_{nm}^{\alpha\beta} Q_{mn}^{\beta\alpha} \rangle .
\end{equation}
With the help of the irreducible part the full correlation function can be restored as follows 
\begin{equation}
P_2^{\alpha\beta}(i\varepsilon_n,i\varepsilon_m)= 4 Z^2 \sgn\varepsilon_n \sgn\varepsilon_m + P_2^{\alpha\beta;{\rm (irr)}}(i\varepsilon_n,i\varepsilon_m).
\end{equation}

Expanding $Q$ to the first order in $W$, we obtain the one-loop contribution,
\begin{gather}
P_{2,1}^{\alpha\beta; {\rm (irr)}}(i\varepsilon_n,i\varepsilon_m) =\mu_2 \langle \tr W_{nm}^{\alpha\beta} W_{mn}^{\beta\alpha}\rangle
=  \frac{16\mu_2}{g} \notag \\
\times \frac{1-\sgn \varepsilon_n\sgn \varepsilon_m}{2}  \int_q \mathcal{D}_q(i|\varepsilon_n|+i|\varepsilon_m|) .
\label{eq:P21:r}
\end{gather} 
Neglecting the energy dependence in the diffusive propagator (for the reasons explained above Eq. \eqref{eq:K2:simp}), we find the following one-loop result for the irreducible part of the operator $K_2$,
\begin{equation}
K_{2,1}^{\rm (irr)}= \frac{\mu_2 t h^\epsilon}{\epsilon} .
\label{eq:K21:irr} 
\end{equation}

\subsubsection{Two-loop renormalization}

Expanding $Q$ to the second order in $W$, we obtain the two-loop contribution as 
\begin{gather}
{}\hspace{-0.3cm} P_{2,2}^{\alpha\beta;{\rm (irr)}}=\frac{1}{4} \sgn \varepsilon_n\sgn \varepsilon_m \Llangle\tr 
(W^2)_{nn}^{\alpha\alpha}\tr (W^2)_{mm}^{\beta\beta}\Rrangle
\notag\\
+ \mu_2 \frac{1+\sgn \varepsilon_n\sgn \varepsilon_m}{8} \Bigl \langle \tr (W^2)_{nm}^{\alpha\beta} (W^2)_{mn}^{\beta\alpha}\Bigr \rangle
\notag \\
+\mu_2 \frac{1-\sgn\varepsilon_n\sgn\varepsilon_m}{2} \Llangle \tr W_{nm}^{\alpha\beta} W_{mn}^{\beta\alpha} 
\begin{bmatrix}
S_{\rm 0}^{(4)}+S_{\rm int}^{(4)}\\
+\frac{1}{2} (S_{\rm int}^{(3)})^2
\end{bmatrix} 
\Rrangle .
\label{eq:K2:2:qw}
\end{gather}
Then, using Eq. \eqref{eq:prop:full}, we obtain
\begin{gather}
\Llangle \tr 
(W^2)_{nn}^{\alpha\alpha}\tr (W^2)_{mm}^{\beta\beta}\Rrangle
= \frac{64}{g^2} \left (\int_{q} \mathcal{D}_q(i|\varepsilon_n|+i|\varepsilon_m|) \right )^2
\notag \\
\to 16 \frac{t^2h^{2\epsilon}}{\epsilon^2} .
\label{eq:trw22}
\end{gather}
In the last line we neglect the energy dependence in the propagators.  We emphasize that this contribution is immune to the interaction. 
Next, we find
\begin{gather}
\Bigl \langle \tr (W^2)_{nm}^{\alpha\beta} (W^2)_{mn}^{\beta\alpha}\Bigr \rangle \! = \!
\frac{32\textsf{v}}{g^2} \!\int\limits_{qp} \mathcal{D}_q(2i|\varepsilon_n|) \mathcal{D}_p(i|\varepsilon_n|+i|\varepsilon_m|) 
\notag \\
- 3 
\frac{128 \pi T \gamma}{g^2 D} \sum_{\varepsilon_k>0} \int\limits_{qp} 
\mathcal{D}_p(i|\varepsilon_m|+i\varepsilon_k)\DD^t_q(i|\varepsilon_n|+i\varepsilon_k)\notag\\
+(n\leftrightarrow m) 
\to  16 \textsf{v} \frac{t^2h^{2\epsilon}}{\epsilon^2}
- 3 \frac{128\gamma}{g^2} J_{101}^0(1+\gamma)
\notag \\
\simeq 16 \textsf{v} \frac{t^2h^{2\epsilon}}{\epsilon^2}
- 48 \frac{t^2h^{2\epsilon}}{\epsilon^2}
 \Bigl [ \ln(1+\gamma)-\frac{\epsilon}{4}\ln^2(1+\gamma)\Bigr] .
\label{eq:trw23}
\end{gather}
Here we introduce the following notation for integral over momenta and frequency,
\begin{gather}
J^\delta_{\nu\mu\eta}(a)= \int_{qp} \int\limits_0^\infty ds \, s^\delta \frac{1}{(p^2+h^2+s)^\nu}\frac{1}{p^2+h^2+as}
\notag \\
\times \frac{1}{(q^2+h^2)^\mu}\frac{1}{(\bm{p}+\bm{q})^2+h^2+s)^\eta} .
\end{gather}
The  integrals $J^\delta_{\nu\mu\eta}$ were computed in Ref. \cite{Burmistrov2015m}.

Using the result \eqref{eq:S4sigma:part:int}, we obtain
\begin{gather}
\Llangle\! \tr [W_{nm}^{\alpha\beta}W_{mn}^{\beta\alpha}] S_{\rm 0}^{(4)} \!\Rrangle
\!=\! - \frac{8 \textsf{v}}{g^2}\! \int\limits_{qp} \mathcal{D}_q(i|\omega_{nm}|)
 \Bigl [ 2 \mathcal{D}_q(i|\omega_{nm}|)
 \notag\\
  +\mathcal{D}_p(i2 |\varepsilon_n|) +  \mathcal{D}_p(i2 |\varepsilon_m|)\Bigr ] 
 +3 
 \frac{32 \pi T \gamma}{g^2 D} \! \int\limits_{qp} \Bigl[ \sum_{\omega_k>|\varepsilon_{n}|} 
 \notag \\
 + \sum_{\omega_k>|\varepsilon_{m}|}\Bigr ]
 \Bigl [
 \mathcal{D}_q(i|\omega_{nm}|)
 + \mathcal{D}_p(i\omega_k) 
 \Bigr ]\mathcal{D}^t_p(i\omega_k)\mathcal{D}_q(i|\omega_{nm}|)
 \notag\\
 \to -4 \textsf{v}  \frac{t^2 h^{2\epsilon}}{\epsilon^2} 
  + 3 \frac{32 \gamma}{g^2}   \Bigl [J^0_{020}(1+\gamma)+ J^0_{110}(1+\gamma) \Bigr ]
  \notag \\
  \simeq -4 \textsf{v}  \frac{t^2 h^{2\epsilon}}{\epsilon^2} 
  + 24 \frac{t^2 h^{2\epsilon}}{\epsilon^2} 
 \Bigl [\ln(1+\gamma)+ \frac{\epsilon\gamma}{2(1+\gamma)} \Bigr ] .
\label{eq:trw24}
\end{gather}

In order to compute the last contribution in Eq. \eqref{eq:K2:2:qw}, we use the following simplification (we note that it is possible due to different replica indices, $\alpha\neq\beta$):
\begin{gather}
S^{(4)}_{\rm int}+\frac{1}{2}(S^{(3)}_{\rm int})^2  \to -  \sum_{\alpha n} \int\limits_{\bm{x}, \bm{x^\prime}}
\Bigl [1-\frac{\gamma |\omega_n|}{D}\mathcal{D}^t_{\bm{x}-\bm{x^\prime}}(i|\omega_n|)\Bigr  ]
\notag \\
\times
\frac{\pi T\Gamma_t }{4}\sum_{\textsf{j}=1}^3 \Tr I^{\alpha}_{n} \sj \Lambda W^2(\bm{x})
\Tr I^{\alpha}_{-n} \sj\Lambda W^2(\bm{x^\prime})
 .
\end{gather}
After tedious but straightforward calculations, we find 
\begin{gather}
{}\hspace{-0.3cm}\Llangle \tr W_{nm}^{\alpha\beta} W_{mn}^{\beta\alpha} \Bigl [S_{\rm int}^{(4)}
+\frac{1}{2} (S_{\rm int}^{(3)})^2\Bigr ]\Rrangle =
-
3\frac{32\pi T\gamma}{g^2 D}  \int\limits_{pq}
\notag \\ \times
\Bigl (\sum_{|\varepsilon_n|>\omega_k}+\sum_{|\varepsilon_m|>\omega_k}\Bigr )\Bigl [ 1 - \frac{\gamma |\omega_k|}{D}\mathcal{D}^t_{\bm{p}+\bm{q}}(i|\omega_k|)  \Bigr ]
\notag \\
\times
\mathcal{D}^2_{p}(i|\varepsilon_n|+i|\varepsilon_m|)
\mathcal{D}_q(i|\varepsilon_n|+i|\varepsilon_m|-i\omega_k)   \notag \\
\to - \frac{96\gamma}{g^2}\Bigl [ J_{020}^0(1)-\gamma J_{021}^1(1+\gamma)\Bigr ]
\simeq
- 12 \gamma\frac{t^2h^{2\epsilon}}{\epsilon^2}  
\notag \\
\times \Bigl [ 
\frac{2\gamma -(2+\gamma)\ln(1+\gamma)}{\gamma^2} 
+\epsilon \frac{(2+\gamma)\ln(1+\gamma)}{\gamma^2}
\notag \\
+ \frac{\epsilon}{1+\gamma} 
+\epsilon \frac{2+\gamma}{\gamma^2} \Bigl (\li2(-\gamma) 
+\frac{1}{4}\ln^2(1+\gamma)\Bigr )
\Bigr ] .
\label{eq:trw25}
\end{gather}
Here $\li2(z)=\sum_{k=1}^\infty z^k/k^2$ denotes the polylogarithm. 
We note that the factor $32$ in the first line of the above equation appears as the result of the following identity for $\textsf{j}=1,2,3$:
\begin{equation}
\sum_{\textsf{j}_{1,2}=0}^3 
\tr (\sj \textsf{s}_{\textsf{j}_1}\textsf{s}_{\textsf{j}_2}) \Bigl [
\tr (\sj \textsf{s}_{\textsf{j}_2}\textsf{s}_{\textsf{j}_1}) 
- \textsf{v}_{\textsf{j}_1}\textsf{v}_{\textsf{j}_2} 
\tr (\sj \textsf{s}_{\textsf{j}_1}\textsf{s}_{\textsf{j}_2})
\Bigr ] =32  .
\end{equation}

Combing the above results, Eqs. \eqref{eq:trw22}--\eqref{eq:trw24} and \eqref{eq:trw25}, we find
\begin{equation}
K_{2,2}^{\rm (irr)} = \Bigl [ \mu_2(\textsf{v}-6\ln(1+\gamma)) + (b_2^{(2)}+\epsilon \mu_2 b_3)\Bigr ] \frac{t^2h^{2\epsilon}}{\epsilon^2},  
\label{eq:K22:irr} 
\end{equation}
where
\begin{gather} 
b_2^{(2)} = 1  -  3 \mu_2 f(\gamma) , \quad 
b_3 = \frac{3}{2} \Bigl \{\frac{1+\gamma}{2\gamma} \ln^2(1+\gamma)
\notag \\
+  \frac{2+\gamma}{\gamma} \Bigl [\li2(-\gamma)+\ln(1+\gamma)\Bigr ]
 \Bigr \}.
\end{gather}

\subsubsection{Anomalous dimension}

Employing the one-loop (see Eq. \eqref{eq:K21:irr}) and two-loop (see Eq. \eqref{eq:K22:irr}) results, we write the operator $K_2$ in the following form 
\begin{equation}
K_2 = Z^2 M_2 K_2[\Lambda] .
\end{equation}
Here $K_2[\Lambda]=1$ is the classical value of $K_2$ and 
\begin{gather}
M_2 = 
1 +Z^{-2} (K_{2,1}^{\rm (irr)}+K_{2,2}^{\rm (irr)})
= 1 + \mu_2 \frac{t h^\epsilon}{\epsilon}+(b_2^{(2)}
\notag \\ +\epsilon \mu_2 b_3) 
\frac{t^2 h^{2\epsilon}}{\epsilon^2}=
1 + \mu_2 \frac{t h^{\prime \epsilon}}{\epsilon}+(b_2^{(2)}+\epsilon\mu_2  \tilde{b}_3)  \frac{t^2 h^{\prime 2\epsilon}}{\epsilon^2} .
\label{eq:K2:int:1}
\end{gather}
where $\tilde{b}_3=b_3+b/2$ with $b$ given by Eq. \eqref{eq:hprime:ren}. We note that in order to determine $M_2$ within the two-loop approximation it is enough to know the factor $Z$ in the one-loop approximation only. That is why we considered irreducible part of $K_2$.

Applying the minimal subtraction scheme to Eq. \eqref{eq:K2:int:1}, we obtain the anomalous dimension of $M_2$ upto the second order in $t$:
\begin{equation}
\eta^{(\mu_2)} = - \frac{d\ln M_2}{d\ell} =  \mu_2 \Bigl [ t + 3 c(\gamma) t^2 \Bigr ]+ O(t^3) .
\label{eq:eta:2}
\end{equation}
Here we introduce the function (cf. Refs. \cite{Burmistrov2013,Burmistrov2015m,Repin2016})
\begin{equation}
c(\gamma) = 2 + \frac{1+\gamma}{2\gamma}\ln^2(1+\gamma)+\frac{2+\gamma}{\gamma}\li2(-\gamma) .
\label{eq:def:c:gamma}
\end{equation}
The finiteness of $\eta_{\mu_2}$ in the limit $\epsilon\to 0$ is guaranteed if the following condition holds:
\begin{equation}
\mu_2 \left (\mu_2-a_1\right ) = 2b_2^{(2)} \quad \Leftrightarrow\quad
\mu_2(\mu_2-1)=2 .
\label{eq:self-cons:2}
\end{equation} 
We emphasize that the self-consistent condition \eqref{eq:self-cons:2} is (i) nonlinear in $\mu_2$ and (ii) independent of the interaction strength $\gamma$. The former implies that it cannot be satisfied by a linear combination of two or more eigenoperators. The later guarantees that the eigenoperators in the absence of interaction remain eigenoperators in the presence of interaction (see more detailed discussion in Sec. \ref{Sec:Final} below).  

Solving Eq. \eqref{eq:self-cons:2} we find two solutions: $\mu_2=2,-1$ in full agreement with the background field renormalization scheme above.  We emphasize that Eq. \eqref{eq:self-cons:2} uniquely determines the value of $\mu_2$. Denoting the corresponding eigenoperator as in the noninteracting case, we find
\begin{gather}
\begin{array}{cc}
\mu_2=-1, \qquad & \eta_{\textsf{(2)}} = - t (1+3c(\gamma) t) +O(t^3) ,\\
 \mu_2=2, \qquad & \eta_{\textsf{(1,1)}} = 2 t (1+3c(\gamma) t)+O(t^3) .
\end{array}
\label{eq:anom:q2}
\end{gather}

\subsection{Operators with three $Q$ matrices}

Now we switch to the eigenoperators with three $Q$ matrices. As above, in order to determine these eigenoperators, we start from the background field renormalization.  

\subsubsection{Background field renormalization}

The operators with three $Q$ matrices are constructed with the help of 
the following Matsubara operator
\begin{gather}
P_3^{\alpha\beta\mu}(i\varepsilon_k,i\varepsilon_n,i\varepsilon_m)=
\tr Q_{kk}^{\alpha\alpha}  \tr Q_{nn}^{\beta\beta} \tr Q_{mm}^{\mu\mu}
+\mu_{2,1} \tr Q_{kk}^{\alpha\alpha}\notag \\\times  \tr Q_{nm}^{\beta\mu} Q_{mn}^{\mu\beta}+\mu_3 \tr Q_{kn}^{\alpha\beta}Q_{nm}^{\beta\mu} Q_{mk}^{\mu\alpha} .
\end{gather}
We start renormalization from the operator with three traces (since the methodology is similar to renormalization of operators with two $Q$ matrices, we present the final results only):
\begin{gather}
R_{1,1,1}[Q]\to Z^3 R_{1,1,1}[\underline{Q}] -3 Y R_{2,1}[\underline{Q}]+3Y R_{\bar{2},1}[\underline{Q}]
\label{eq:BGFR:3Q:111}
\end{gather}
Here the `bar' sign on the index denotes the sign change of one Matsubara frequency, e.g.  
\begin{equation}
R_{\bar{2},1}[Q]=\tr Q_{k,-n}^{\alpha\beta}  Q_{-n,k}^{\beta\alpha} \tr Q_{mm}^{\mu\mu} .
\end{equation}

Next, the operator with two traces renormalizes as follows
\begin{gather}
R_{2,1}[Q]\to Z^3 R_{2,1}[\underline{Q}] - Y R_{2,1}[\underline{Q}]+Y R_{\bar{2},1}[\underline{Q}]
-2 Y R_{3}[\underline{Q}]\notag \\
+2Y R_{\bar{3}}[\underline{Q}] .
\label{eq:BGFR:3Q:21}
\end{gather}
Finally, the operator with single trace transforms in the following way
\begin{gather}
R_3[Q]\to Z^3 R_3[\underline{Q}] - 3 Y R_{2,1}[\underline{Q}]+3 Y R_{\bar{3}}[\underline{Q}] .
\label{eq:BGFR:3Q:3}
\end{gather}
Using the above results, Eqs. \eqref{eq:BGFR:3Q:111} -- \eqref{eq:BGFR:3Q:3}, we write  
\begin{gather}
K_3[Q] \!\to\! Z^3 K_3[\underline{Q}]\!-\!\mu_{2,1} Y O_{1,1,1}[\underline{Q}]
-(4\mu_{2,1}+3\mu_3) Y O_3[\underline{Q}]\notag \\
- (\mu_{2,1}+3(2+\mu_3)) Y O_{2,1}[\underline{Q}]
\simeq Z^3 (1-\mu_{2,1} Y) \Bigl [ O_{1,1,1} \notag \\
 +(\mu_{2,1}+(\mu_{2,1}^2-\mu_{2,1}-6-3\mu_3)Y)O_{2,1}
 \notag \\
 \hspace{0.6cm}+(\mu_3+(\mu_3\mu_{2,1}-4\mu_{2,1}-3\mu_3)Y)O_3\Bigr] .
\end{gather}
As one can see from the above expression, in order $K_3$ to be the eigenoperator under the renormalization group transformation, the coefficients $\mu_{2,1}$ and $\mu_3$ have to satisfy the following system of equations, 
\begin{equation}
\mu_{2,1}^2-\mu_{2,1}=6+3\mu_3 , 
\quad
\mu_3(\mu_{2,1}-3) = 4\mu_{2,1} .
\label{eq:eq:for:coef:3}
\end{equation}
This system of equations reduces to the cubic equation for $\mu_{2,1}$. There are three pairs of solutions of Eqs. \eqref{eq:eq:for:coef:3},
\begin{equation}
\begin{array}{ccccc}
\mu_{2,1} & = & -3, & 1, & 6, \\
\mu_3 & = & 2, & -2, & 8,
\end{array}
\label{eq:set:of:mu:3}
\end{equation}
corresponding to three eigenoperators. Again we note that interaction does not affect one-loop renormalization of the operators without derivatives.   

We note in passing that Eqs. \eqref{eq:eq:for:coef:3} can be cast in the form of the eigenvalue problem:
\begin{gather}
\mathcal{M}_3^\textsf{T}
\begin{pmatrix}
1\\
\mu_{2,1}\\
\mu_3
\end{pmatrix}
=\mu_{2,1} \begin{pmatrix}
1\\
\mu_{2,1}\\
\mu_3
\end{pmatrix}, \quad \mathcal{M}_3=
\begin{pmatrix}
0 & 6 & 0\\
1 & 1 & 4 \\
0 & 3 & 3
\end{pmatrix} .
\label{eq:M3:def}
\end{gather}
The matrix $\mathcal{M}_3$ is nothing but the matrix describing the mixing of operators $O_{1,1,1}$, $O_{2,1}$, and $O_3$ under the action of the renormalization group. We emphasize that the values of $\mu_{2,1}$ in Eq. \eqref{eq:set:of:mu:3} coincides with the eigenvalues of the mixing matrix $\mathcal{M}_3$.

\subsubsection{Two-loop renormalization}

As for the operators with two $Q$ matrices it is convenient to single out the irreducible part from $K_3$ by subtracting $\langle Q\rangle$ from $Q$. Using explicit knowledge of eigenoperators with two $Q$ matrices, we find
\begin{gather}
K_3 = K_3^{\rm (irr)} -2Z^3 + Z^3 \Bigl (2 M_{\textsf{(2)}}  + M_{\textsf{(1,1)}}\Bigr )
\notag \\
+ \frac{\mu_{2,1}}{3} Z^3 \Bigl (M_{\textsf{(1,1)}} -M_{\textsf{(2)}}\Bigr ) .
\label{eq:K3:int:01}
\end{gather}

There is no one-loop contribution to $K_3^{\rm (irr)}$. At the two-loop approximation we find the following expression for the irreducible part of the  corresponding operator $P_{3}^{\alpha\beta\gamma}(i\varepsilon_k,i\varepsilon_n,i\varepsilon_m)$:
\begin{gather}
P_{3}^{\alpha\beta\gamma;{\rm (irr)}}(i\varepsilon_k,i\varepsilon_n,i\varepsilon_m) \simeq 
-\frac{3}{2} \mu_3 \sgn\varepsilon_k \frac{1+\sgn\varepsilon_k\sgn\varepsilon_n}{2}
\notag \\
\times
\frac{1-\sgn\varepsilon_k\sgn\varepsilon_m}{2}\langle\tr (W^2)_{kn}^{\alpha\beta} W_{nm}^{\beta\gamma} W_{mk}^{\gamma\alpha} \rangle
\notag \\
= -\frac{3}{2} \mu_3 \sgn\varepsilon_k \frac{1+\sgn\varepsilon_k\sgn\varepsilon_n}{2}\frac{1-\sgn\varepsilon_k\sgn\varepsilon_m}{2} \notag \\
\times \frac{128}{g^2} \int_{qp}\mathcal{D}_q(i|\varepsilon_n|+i|\varepsilon_m|)
\mathcal{D}_p(i|\varepsilon_k|+i|\varepsilon_m|)  .
\end{gather}
Hence, we obtain the following two-loop result for the irreducible part of $K_3$: 
\begin{equation}
K_3^{\rm (irr)} = \frac{3 \mu_3}{2} \frac{t^2h^{2\epsilon}}{\epsilon^2} . 
\label{eq:K3:int:1}
\end{equation}
We emphasize that it is independent of the interaction.

\subsubsection{Anomalous dimension}

Using the results \eqref{eq:K3:int:01} and \eqref{eq:K3:int:1}, we derive the following expression  
\begin{gather}
K_3 = Z^3 M_3 K_3[\Lambda] ,
\end{gather}
where
\begin{gather}
M_3=1 + \frac{\mu_{2,1} t h^{\prime \epsilon}}{\epsilon}
+(b_2^{(3)}+\epsilon \mu_{2,1} \tilde{b}_3)\frac{t^2 h^{\prime 2\epsilon}}{\epsilon^2} ,\notag \\
b_2^{(3)} = 3 +\frac{3\mu_{3}}{2}+\frac{(1-a_1)\mu_{2,1}}{2} .
\end{gather}
The anomalous dimension of $M_3$ is given (in the limit $\epsilon\to 0$) as
\begin{gather}
\eta^{(\mu_{2,1})} = - \frac{d\ln M_{3}}{d\ell} = \mu_{2,1} t \bigl [1+3c(\gamma)t\bigr ] +  O(t^3) ,
\end{gather}
provided the following relation holds 
\begin{equation}
(\mu_{2,1}-a_1)\mu_{2,1}=2b_2^{(3)} \, \Rightarrow \, \mu_{2,1}^2-\mu_{2,1} = 3\mu_3+6 . 
\label{eq:rel:cons:3}
\end{equation}
As in the case of operators with two $Q$ matrices, the consistency condition is independent of the interaction parameter $\gamma$. However, it does not determine the eigenoperators uniquely. Since Eq. \eqref{eq:rel:cons:3} is nothing but the first equation in the system of equations \eqref{eq:eq:for:coef:3}, three sets of values of $\mu_{2,1}$ and $\mu_3$ found within background field renormalization, cf. Eq. \eqref{eq:set:of:mu:3}, do satisfy Eq. \eqref{eq:rel:cons:3}. 
If we want to find the values of $\mu_{2,1}$ and $\mu_3$ from the direct computation of the averages of the operators, we would need to go to the three-loop order. 

All in all, we find the following anomalous dimensions of three eigenoperators with three $Q$ matrices:
\begin{equation}
\begin{array}{ccc}
\mu_{2,1} = -3,  & \mu_3=2,\qquad & \eta_{\textsf{(3)}} = - 3t (1+3c(\gamma)t),\\
\mu_{2,1}=1, & \mu_3=-2,\qquad & \eta_{\textsf{(2,1)}} = t (1+3c(\gamma)t), \\
\mu_{2,1}=6, & \mu_3=8,\qquad  & \eta_{\textsf{(1,1,1)}} = 6 t (1+3c(\gamma)t) .
\end{array}
\label{eq:anom:q3}
\end{equation}

\begin{widetext}

\subsection{Operators with four $Q$ matrices} 

\subsubsection{Background field renormalization}

The operators with four $Q$ matrices are constructed with the help of 
the following Matsubara operator
\begin{gather}
P_4^{\alpha\beta\mu\nu}(i\varepsilon_k,i\varepsilon_n,i\varepsilon_m,i\varepsilon_l)=
\tr Q_{kk}^{\alpha\alpha}  \tr Q_{nn}^{\beta\beta} \tr Q_{mm}^{\mu\mu}\tr Q_{ll}^{\nu\nu}
+\mu_{2,1,1} \tr Q_{kn}^{\alpha\beta} Q_{nk}^{\beta\alpha} \tr Q_{mm}^{\mu\mu}
\tr Q_{ll}^{\nu\nu} 
{+}\mu_{3,1} \tr Q_{kn}^{\alpha\beta}Q_{nm}^{\beta\mu} Q_{mk}^{\mu\alpha}
\tr Q_{ll}^{\nu\nu} \notag \\
+\mu_{2,2} \tr Q_{kn}^{\alpha\beta} Q_{nk}^{\beta\alpha} \tr Q_{ml}^{\mu\nu} Q_{lm}^{\nu\mu} 
+ \mu_4 \tr Q_{kn}^{\alpha\beta}Q_{nm}^{\beta\mu} Q_{ml}^{\mu\nu} Q_{lk}^{\nu\alpha} .
\end{gather}
Performing background field renormalization in the same way as above, we find the following results
\begin{subequations}
\begin{align}
&R_{1,1,1,1}[Q] \to Z^4 R_{1,1,1,1}[\underline{Q}] -6 Y R_{2,1,1}[\underline{Q}] +6 Y R_{\bar{2},1,1}[\underline{Q}] ,
\label{eq:R1111}
 \\
&R_{2,1,1}[Q] \to Z^4 R_{2,1,1}[\underline{Q}] -Y R_{1,1,1,1}[\underline{Q}]
+Y R_{\bar{2},1,1}[\underline{Q}] -Y R_{2,2}[\underline{Q}]+Y R_{\bar{2},2}[\underline{Q}]-  4 Y R_{3,1}[\underline{Q}]+4 Y R_{\bar{3},1}[\underline{Q}],
\label{eq:R211}
\\
&R_{3,1}[Q] \to Z^4 R_{3,1}[\underline{Q}] -3 Y R_{2,1,1}[\underline{Q}]
+3 Y R_{\bar{3},1}[\underline{Q}]  - 3Y R_{4}[\underline{Q}]+3 Y R_{\bar{4}}[\underline{Q}],
\label{eq:R31}
\\
&R_{2,2}[Q] \to Z^4 R_{2,2}[\underline{Q}]-2 Y R_{2,1,1}[\underline{Q}] +2 Y R_{\bar{2},2}\underline{Q}]\ -4Y R_{4}[\underline{Q}]+4 Y R_{\bar{4}}[\underline{Q}],
\label{eq:R22}\\
&R_4[Q] \to Z^4 R_4[\underline{Q}]-4 Y R_{3,1}[\underline{Q}]-2 Y R_{2,2}[\underline{Q}]+6 Y R_{\bar{4}}[\underline{Q}] .
\label{eq:R4}
\end{align}
\end{subequations}
Using the above expressions, we obtain
\begin{gather}
K_4[Q]\to Z^4 K_4[\underline{Q}] 
- \mu_{2,1,1}  Y O_{1,1,1,1} - (12+\mu_{2,1,1}+3\mu_{3,1}
+2\mu_{2,2}) Y O_{2,1,1}
- (8\mu_{2,1,1}+3\mu_{3,1}+4\mu_4) Y O_{3,1}
-(2\mu_{2,1,1}\notag \\
+2\mu_{2,2}+2 \mu_4) Y O_{2,2}
-(6 \mu_{3,1}+8 \mu_{2,2} +6 \mu_4) Y O_4 \simeq Z^4 (1- \mu_{2,1,1}  Y) \Bigl [ O_{1,1,1,1} 
+(\mu_{2,1,1}+(\mu_{2,1,1}^2-12-\mu_{2,1,1}-3\mu_{3,1}
\notag \\
-2\mu_{2,2}) Y)O_{2,1,1}
+(\mu_{3,1} +(\mu_{3,1}\mu_{2,1,1}-8\mu_{2,1,1}-3\mu_{3,1}-4\mu_4) Y)O_{3,1}
+(\mu_{2,2}+(\mu_{2,2}(\mu_{2,1,1}-2)-2\mu_{2,1,1} -2 \mu_4) Y) O_{2,2}
\notag \\
+(\mu_4 +(\mu_4\mu_{2,1,1}-6 \mu_{3,1}-8 \mu_{2,2} -6 \mu_4) Y) O_4 
\Bigr ] .
\end{gather}
\end{widetext}
In order $K_3$ to be the eigenoperator under the renormalization group transformation, the coefficients $\mu_{2,1,1}, \mu_{3,1}, \mu_{2,2}$ and $\mu_4$ have to satisfy the following system of equations,
\begin{subequations}
\begin{align}
\mu_{2,1,1}^2 & =12+\mu_{2,1,1}+3\mu_{3,1}+2\mu_{2,2} ,\label{eq:4.1} \\
\mu_{3,1}\mu_{2,1,1}& = 8\mu_{2,1,1}+3\mu_{3,1}+4\mu_4 ,\label{eq:4.2}\\
\mu_{2,2} \mu_{2,1,1}& =  2\mu_{2,1,1} + 2\mu_{2,2}+ 2 \mu_4 , \label{eq:4.3}\\
\mu_4\mu_{2,1,1}&=6 \mu_{3,1}+8 \mu_{2,2}+6 \mu_4 .\label{eq:4.4}
\end{align}
\end{subequations}
This system of equations reduces to the fifth degree algebraic equation for $\mu_{2,1,1}$. There are five sets of solutions of Eqs. \eqref{eq:4.1}-\eqref{eq:4.4},
\begin{equation}
\begin{array}{ccccccc}
\mu_{2,1,1} & = & 12, & -6, & 5, & 2, & -1, \\  
\mu_{3,1} & = & 32, & 8, & 4, & -8, & -2, \\
\mu_{2,2} & = & 12, & 3, & -2, & 7, & -2, \\
\mu_{4} & = & 48, & -6, & -8, & -2, & 4,
\end{array}
\label{eq:set:of:mu:4}
\end{equation}
corresponding to five eigenoperators. Since interaction does not affect one-loop renormalization of the operators without derivatives these operators are exactly the same as in the absence of interaction.   

As in the case of $q=3$ operators, Eqs. \eqref{eq:4.1}-\eqref{eq:4.4} can be cast in the form of the eigenvalue problem:
\begin{gather}
\mathcal{M}_4^\textsf{T}
\begin{pmatrix}
1\\
\mu_{2,1,1}\\
\mu_{3,1}\\
\mu_{2,2}\\
\mu_4
\end{pmatrix}
=\mu_{2,1,1} \begin{pmatrix}
1\\
\mu_{2,1,1}\\
\mu_{3,1}\\
\mu_{2,2}\\
\mu_4
\end{pmatrix}, \notag \\ 
\mathcal{M}_4=
\begin{pmatrix}
0 & 12 & 0 & 0 & 0 \\
1 & 1 & 8 & 2 & 0 \\
0 & 3 & 3 & 0 & 6 \\
0 & 2 & 0 &  2 &  8\\
0 & 0 & 4 & 2 & 6
\end{pmatrix} .
\label{eq:M4:def}
\end{gather}
The matrix $\mathcal{M}_4$ describes the mixing of operators $O_{1,1,1,1}$, $O_{2,1,1}$, $O_{3,1}$, $O_{2,2}$, and $O_4$ under the action of the renormalization group. We note that the magnitudes of $\mu_{2,1,1}$ in Eq. \eqref{eq:set:of:mu:4} coincides with the eigenvalues of the mixing matrix $\mathcal{M}_4$.

\subsubsection{Two-loop renormalization}

With the help of explicit knowledge of eigenoperators with two and three $Q$ matrices, we express $K_4$ as a linear combination of reducible and irreducible parts,
\begin{gather}
K_4 = Z^4 \Bigl (3 - 2 \bigl [2M_\textsf{(2)}{+}M_\textsf{(1,1)}\bigr ]+\frac{\mu_{2,1,1}}{3}
\bigl [M_\textsf{(2)}{-}M_\textsf{(1,1)}\bigr ] \notag \\
{+}\frac{4}{15} \bigl [M_\textsf{(1,1,1)}{+}9 M_\textsf{(2,1)}{+}5 M_\textsf{(3)}\bigr ] {+}\frac{\mu_{2,1,1}}{15}
\bigl [2M_\textsf{(1,1,1)}{+}3 M_\textsf{(2,1)}
\notag \\
{-}5 M_\textsf{(3)}\bigr ]
{+}\frac{\mu_{3,1}}{60} \bigl [4M_\textsf{(1,1,1)}{-}9 M_\textsf{(2,1)}{+}5 M_\textsf{(3)}\bigr ]\Bigr )
{+}K_4^{\rm (irr)} .
\label{eq:K4:int:01}
\end{gather}
There is no one-loop contribution to $K_4^{\rm (irr)}$. In the next order we obtain the following expression for the irreducible part of the  corresponding Matsubara operator:
\begin{gather}
P_4^{\alpha\beta\mu\nu;{\rm (irr)}}(i\varepsilon_k,i\varepsilon_n,i\varepsilon_m,i\varepsilon_l)  \simeq 
\mu_{2,2} \langle 
 \tr W_{kn}^{\alpha\beta} W^{\beta\alpha}_{nk} 
 \tr W_{ml}^{\mu\nu} \notag \\ \times W^{\nu\mu}_{lm} \rangle = \frac{256 \mu_{2,2}}{g^2} \frac{1{-}\sgn \varepsilon_k\sgn\varepsilon_n}{2} \frac{1{-}\sgn \varepsilon_m\sgn\varepsilon_l}{2}\notag \\
\times \int\limits_{pq} \mathcal{D}_p(i|\varepsilon_k|+i|\varepsilon_n|) 
 \mathcal{D}_q(i|\varepsilon_m|+i|\varepsilon_l|) .
\end{gather}
Hence, we derive the following two-loop result: 
\begin{gather}
K_4^{\rm (irr)} = \mu_{2,2} \frac{t^2 h^{2\epsilon}}{\epsilon^2} .
\label{eq:K4:int:1}
\end{gather}
We emphasize its independence of the interaction.

\subsubsection{Anomalous dimension}

With the help of the results \eqref{eq:K4:int:01} and \eqref{eq:K4:int:1}, we derive the following expression for the renormalized operator $K_4$,
\begin{gather}
K_4 = Z^4 M_4 K_4[\Lambda] ,
\end{gather}
where
\begin{gather}
M_4=1 + \frac{\mu_{2,1,1} t h^{\prime \epsilon}}{\epsilon}
+(b_2^{(4)}+\epsilon \mu_{2,1,1} \tilde{b}_3)\frac{t^2 h^{\prime 2\epsilon}}{\epsilon^2} ,\notag \\
b_2^{(4)} = 6 +\frac{3\mu_{3,1}}{2}+\mu_{2,2}+\frac{(1-a_1)\mu_{2,1,1}}{2} .
\end{gather}
The anomalous dimension of $M_4$ is given as
\begin{gather}
\eta^{(\mu_{2,1,1})} = - \frac{d\ln M_{4}}{d\ell} = \mu_{2,1,1} t [1+3c(\gamma)t] +  O(t^3) ,
\end{gather}
provided the following relation holds 
\begin{equation}
(\mu_{2,1,1}-a_1)\mu_{2,1,1}=2b_2^{(3)}  .
\end{equation}
As one can check, $a_1$ in this equation drops from both sides and the equation becomes
\begin{equation}
\mu_{2,1,1}^2-\mu_{2,1,1} = 12+3\mu_{3,1}+2\mu_{2,2} . 
\label{eq:rel:cons:4}
\end{equation}
As in the case of operators with two and three $Q$ matrices, the consistency condition is independent of the interaction parameter $\gamma$. However, it does not determine the eigenoperators uniquely. Eq. \eqref{eq:rel:cons:4} coincides with Eq. \eqref{eq:4.1}. Therefore, all five sets of parameters $\{\mu_{2,1,1}, \mu_{3,1}, \mu_{2,2}, \mu_{4}\}$, cf. Eq. \eqref{eq:set:of:mu:4}, do satisfy Eq. \eqref{eq:rel:cons:4}.  In order to find the their values from computation of quantum corrections one has to consider to higher orders in the loop expansion. 

Now we list  the following anomalous dimensions of five eigenoperators with four $Q$ matrices:
\begin{align}
\{-6, 8, 3, -6\}, & \quad \eta_{\textsf{(4)}} = - 6t (1+3c(\gamma)t), \notag \\
\{-1, -2, -2, 4\} , & \quad \eta_{\textsf{(3,1)}} = - t (1+3c(\gamma)t), \notag \\
\{2, -8, 7, -2\} , & \quad \eta_{\textsf{(2,2)}} = 2t (1+3c(\gamma)t),  \label{eq:anom:q4} \\
\{
5, 4, -2, -8\} , & \quad \eta_{\textsf{(2,1,1)}} = 5t (1+3c(\gamma)t), \notag \\
\{12, 32, 12, 48\} , & \quad \eta_{\textsf{(1,1,1,1)}} = 12t (1+3c(\gamma)t). \notag 
\end{align}
We note that numbers in brackets indicate the set $\{\mu_{2,1,1}, \mu_{3,1}, \mu_{2,2}, \mu_{4}\}$ for a given eigenoperator.

\subsection{Operators with $q>4$}

Now we consider renormalization of the eigenoperators with the number of the $Q$ matrices which is larger than four, $q>4$. We shall benefit from restriction to the two-loop approximation. The crucial simplification is that the irreducible part of the operator ${K}_q$ with $q>4$ vanishes within  the two-loop approximation. Thus the average of the operator ${K}_q$ is fully determined by the reducible part. As we shall see below, the latter 
can be fully expressed as a linear combination of pure scaling operators with the number of the $Q$ matrices less or equal to four. 

At first, we employ the following relations which are valid within two-loop approximation only, 
\begin{gather}
\langle R_{\scriptsize\underbrace{1,\dots,1}_{q}} \rangle {\simeq} 
R_{\scriptsize\underbrace{{1,\dots,1}}_{q}}[\langle Q\rangle ] {+}\frac{q(q{-}1)}{2}
R_{\scriptsize\underbrace{\text{1,\dots,1}}_{q-2}}[\langle Q\rangle ]\langle R_{1,1}[\delta Q]\rangle  ,
\notag \\
\langle R_{\scriptsize2,\underbrace{1,\dots,1}_{q-2}} \rangle {\simeq}
 R_{\scriptsize\underbrace{1,\dots,1}_{q-2}}[\langle Q\rangle] 
 \langle R_2[Q] \rangle {+} (q-2) R_{\scriptsize\underbrace{1,\dots,1}_{q-3}}[\langle Q\rangle] \notag\\
 \times \langle R_{2,1}[\delta Q]\rangle .
\end{gather}
Here, for a sake of brevity, we introduce the notation $\delta Q=Q-\langle Q\rangle$.\footnote{It should not be confused with $\overline{\delta Q}=Q-\Lambda$ introduced previously.} 

Next, we use the following approximate relations
\begin{align}
\langle R_{\scriptsize3,\underbrace{1,\dots,1}_{q-3}} \rangle & \simeq
R_{\scriptsize\underbrace{{1,\dots,1}}_{q-3}}[\langle Q\rangle ] \langle R_{3}[Q]\rangle,  \notag \\
\langle R_{\scriptsize2,2\underbrace{1,\dots,1}_{q-4}} \rangle & \simeq
R_{\scriptsize\underbrace{{1,\dots,1}}_{q-4}}[\langle Q\rangle ] \langle R_{2,2}[Q]\rangle , \\
\langle R_{\scriptsize4,\underbrace{1,\dots,1}_{q-4}} \rangle & \simeq
R_{\scriptsize\underbrace{{1,\dots,1}}_{q-4}}[\langle Q\rangle ] \langle R_{4}[Q]\rangle \notag
\end{align}

For the same reason as the absence of the contribution to the irreducible part of the operator $K_q$ within the two-loop approximation, the averages of more involved operators $R_{k_1,\dots,k_s}$ can be approximated by $R_{k_1,\dots,k_s}[\langle Q\rangle]$.  Next we express $R_{1,\dots,1}[\langle Q\rangle]$ in terms of $Z$. The averages $\langle R_{k_1,\dots,k_s}[Q]\rangle$ with $k_1+\dots+k_s\leqslant 4$ can be written in terms of the previously found eigenoperators.  
Then with the help of  Eqs. \eqref{eq:anom:q2},  \eqref{eq:anom:q3}, and \eqref{eq:anom:q4}, we find 
\begin{widetext}
\begin{gather}
	K_q \simeq Z^q \Biggl \{ 
	1 +\frac{q(q-1)}{6} \bigl  ( M_\textsf{(1,1)}+ 2 M_\textsf{(2)} - 3  \bigr ) + \frac{\mu_{2,1,\dots,1}}{3}  \bigl ( (q-3)(M_\textsf{(2)}-M_\textsf{(1,1)})
	+\frac{(q-2)}{10}(2 M_\textsf{(1,1,1)}+3M_\textsf{(2,1)}-5 M_\textsf{(3)})  \bigr )
	\notag \\
+ \frac{\mu_{3,1,\dots,1}}{60} \bigl (4M_\textsf{(1,1,1)} -9M_\textsf{(2,1)}+5 M_\textsf{(3)} \bigr ) 
	+ \mu_{2,2,1,\dots,1}  \Bigl (\frac{1}{105}M_\textsf{(1,1,1,1)}-\frac{2}{63} M_\textsf{(2,1,1)}+\frac{7}{90} M_\textsf{(2,2)} - \frac{4}{45} M_\textsf{(3,1)} + \frac{1}{30} M_\textsf{(4)}  \Bigr) 
\notag \\
	+ \mu_{4,1,\dots,1}  \Bigl (\frac{1}{105}M_\textsf{(1,1,1,1)}-\frac{2}{63} M_\textsf{(2,1,1)}-\frac{1}{180} M_\textsf{(2,2)} + \frac{2}{45}M_\textsf{(2,1)} - \frac{1}{60} M_\textsf{(4)}  \Bigr ) \Biggr \} .
	\label{eq:Kq:app}
\end{gather}
\end{widetext}
This lengthy expression can be written in a standard form 
\begin{gather}
K_q = Z^q M_q K_q[\Lambda] ,
\end{gather}
where
\begin{gather}
M_q=1 + \frac{\mu_{2,1,\dots,1} t h^{\prime \epsilon}}{\epsilon}
+(b_2^{(q)}+\epsilon \mu_{2,1,\dots,1} \tilde{b}_3)\frac{t^2 h^{\prime 2\epsilon}}{\epsilon^2} ,\notag \\
b_2^{(q)} {=} \frac{q(q-1){+}3\mu_{3,1,\dots,1}{+}(1{-}a_1)\mu_{2,1,\dots,1}}{2} {+}\mu_{2,2,1,\dots,1} .
\end{gather}
Thus the anomalous dimension of $M_q$ is given as
\begin{gather}
\eta^{(\mu_{2,1,\dots,1})} = - \frac{d\ln M_{q}}{d\ell} = \mu_{2,1,\dots,1} t [1+3c(\gamma)t] +  O(t^3) ,
\label{eq:eta:final:q}
\end{gather}
provided the following relation holds 
\begin{equation}
(\mu_{2,1,\dots,1}-a_1)\mu_{2,1,\dots,1}=2b_2^{(q)}   .
\label{eq:rel:f}
\end{equation} 
As it was for operators with $q\leqslant 4$, the interaction terms drop from both sides of Eq. \eqref{eq:rel:f} and it reduces to 
\begin{equation}
\mu_{2,1,\dots,1}^2-\mu_{2,1,\dots,1} = q(q-1)+3\mu_{3,1,\dots,1}+2\mu_{2,2,1,\dots,1} . 
\label{eq:rel:cons:q}
\end{equation}
Therefore, although we do not know explicit expressions for the coefficients $\mu_{k_1,\dots,k_s}$ for $q>4$, we can state that the form of eigenoperators remains the same in spite of the presence of interaction. 

Armed by recent advances in description of the generalized multifractality at the spin quantum Hall transition in the absence of interaction \cite{Karcher2021}, we find that the anomalous dimension of the eigenoperator $M_{\textsf{(k}_\textsf{1},\dots,\textsf{k}_\textsf{s}\textsf{)}}$ is determined within the two loop approximation by the coefficient 
\begin{equation}
\mu_{2,1,\dots,1} = \frac{1}{2} \sum_{\textsf{j}=1}^\textsf{s} \textsf{k}_\textsf{j}(-\textsf{c}_\textsf{j}-\textsf{v}-\textsf{k}_\textsf{j}), \quad \textsf{c}_\textsf{j}=1-4 \textsf{j} .
\label{eq:eta:final:q:2}
\end{equation} 
Here we remind $\textsf{v}=2$.
For example, for the eigen operator $M_{\textsf{(3,1)}}$ we have $\mu_{2,1,1}= [3(1-3)+1(5-1)]/2=-1$ in accordance with Eq. \eqref{eq:anom:q4}.

\section{Discussions and conclusions\label{Sec:Final}}

\subsection{Relations to other symmetry classes}

The results reported in this paper demonstrate that for the symmetry class C in the presence of interaction the eigenoperators with respect to the renormalization group flow can be constructed in the same way as for the noninteracting case. The interaction affects their anomalous dimensions starting from the two-loop order in the expansion in series in dimensionless inverse spin conductance $t$, cf. Eqs. \eqref{eq:eta:final:q} and \eqref{eq:eta:final:q:2}. At the two-loop order the effect of interaction is described by the universal function $c(\gamma)$, cf. Eq. \eqref{eq:def:c:gamma}. We note that the very same function appeared in a similar problem for standard Wigner-Dyson classes \cite{Burmistrov2013,Burmistrov2015m,Repin2016}. It would be interesting to extend the analysis presented in this work to the other two superconducting classes, CI and DIII, that allow interaction within the Finkel'stein {\NLSM}. 

In the absence of interaction in the class C, the eigenoperators written in terms of $Q$ matrices are expressed as disorder averages of the proper combinations of wave functions \cite{Karcher2021}. In the presence of interaction, one can translate the eigenoperators to the proper combinations of the single-particle Green's functions as it was done for class AI in Ref. \cite{Repin2016}. 
 
\subsection{The role of topology} 
 
The nonlinear sigma model for class C allows the presence of the topological $\theta$-term. Although this topological term does not affect the classification of the eigenoperators it contributes to their anomalous dimensions (for bilinear in $Q$ operators for class A see Ref. \cite{Pruisken2005} for details). Also the topological term changes the renormalization group equations for $t$ and $\gamma$ (for class A see Ref. \cite{Pruisken2007} for details). Therefore, generically in the presence of interaction the renormalization for the class C is described by  three-parameter flow in the plane of $t$ (disorder), $\gamma$ (interaction), and $\theta$ (topological angle). The effect of the topological term on renormalization group flow and anomalous dimensions of eigenoperators will be studied elsewhere.   

\subsection{Implications for Weyl symmetry}

At a fixed point of the renormalization group flow corresponding to the spin quantum  
Hall transition, the scaling with the system size $L$ of an eigenoperator characterized by the Young tableau $\lambda=\textsf{(k}_\textsf{1},\dots,\textsf{k}_\textsf{s}\textsf{)}$ (with $\sum_{\textsf{j}=1}^\textsf{s} \textsf{k}_\textsf{j} =q$) is given as
\begin{equation}
K_{\lambda}\sim L^{-x_{\lambda}}, \quad x_{\lambda}= q x_\textsf{(1)}+\Delta_{\lambda} .
\label{eq:scaling:OP}
\end{equation}
Here the exponent $x_\textsf{(1)}$ describes the scaling of the disorder averaged {\DOS} and is equal to the magnitude of $\eta_\textsf{(1)}$ at the fixed point, $x_\textsf{(1)}=\eta_\textsf{(1)}^*$. Similarly, the exponent $\Delta_{\lambda}$ describes the scaling of the operator $M_\lambda$ and coincides with its anomalous dimension at the fixed point, $\Delta_{\lambda}=\eta_\lambda^*$. We note in passing that Eq. \eqref{eq:scaling:OP} states explicitly that the eigenoperators are just the pure scaling operators.

In the absence of interaction, the generalized multifractal dimensions $x_\lambda$ are known to obey symmetry relations as consequence of Weyl-group invariance  \cite{Gruzberg2013}.  The exponents $x_\lambda$ are the same for the eigenoperators related by the following symmetry operations: reflection, $\textsf{k}_\textsf{j}\to -\textsf{c}_\textsf{j} -\textsf{k}_\textsf{j}$, and permutation of some pair, $\textsf{k}_\textsf{j/i}\to \textsf{k}_\textsf{i/j}+(\textsf{c}_\textsf{i/j}-\textsf{c}_\textsf{j/i})/2$.
For example, reflection symmetry implies that $x_\textsf{(q)}=x_\textsf{(3-q)}$, i.e., in particular,  $x_\textsf{(1)}=x_\textsf{(2)}$ and $x_\textsf{(3)}=0$. Of course, the one-loop results for the anomalous dimensions (in the absence of interaction) is consistent with the symmetry relations. 

We emphasize that the presence of interaction seems to break the symmetry relations between exponents. It can be seen already at one-loop order. The interaction affects $\eta_\textsf{(1)}$ but leaves anomalous exponents for the eigenoperators with $q\geqslant 2$ intact. We consider the transition in $d=2+\epsilon$ dimensions. Then according to the renormalization group equations \eqref{eq:RG:one-loop}, there is a line of fixed points at $t_*=\epsilon/(1+6f(\gamma))$ and arbitrary $\gamma$. The generalized multifractal exponents becomes (to the order $\epsilon$)
\begin{equation}
x_\lambda =  \frac{\epsilon}{2[1+6f(\gamma)]} \sum_{\textsf{j}=1}^\textsf{s} \textsf{k}_\textsf{j}(-\textsf{c}_\textsf{j}-3\ln(1+\gamma)-\textsf{k}_\textsf{j}) .
\label{eq:xLambda:2+e}
\end{equation} 
As one can see, the above expression is inconsistent with Weyl symmetry for $\gamma\neq 0$. Although we have no access to values of generalized multifractal dimensions at the spin quantum Hall transition in $d=2$, we do not see why the Weyl symmetry relations should hold in the presence of interaction. It would be challenging to compute $x_\lambda$ numerically for $\gamma\neq 0$. 

It is worth mentioning that in standard Wigner-Dyson classes interaction remains intact the Weyl-group symmetry relations within the second order expansion in $\epsilon$, provided the relations are formulated for  anomalous dimensions $\Delta_\lambda$ rather than exponents $x_\lambda$. It can be done since $x_\textsf{(1)}\equiv 0$ in the absence of interaction for classes A, AI, and AII. In contrast, our results suggest breakdown of these symmetry relations in class C by interaction.

In the absence of interaction some of generalized multifractal exponents $x_\lambda>0$ at the spin quantum Hall transition can be computed analytically by exact mapping of corresponding observables to hull operators in classical percolation \cite{Gruzberg1999,Beamond2002,Mirlin2003,Evers2003,Subramaniam2008}. Recently, a list of exact analytical results for $x_\lambda$ has been significantly extended \cite{Karcher2022}. It would be interesting to investigate mapping to the percolation in the presence of interaction.

\subsection{Renormalization of interaction at the spin quantum Hall transition}

One-loop renormalization group \eqref{eq:RG:one-loop} predicts that the interaction $\gamma$ does not flow with the length scale. We mention that this fact is intimately related with the existence of Weyl symmetry in the absence of interaction. For weak interaction, $|\gamma|\ll 1$, one can write down the renormalization group equation for $\gamma$ at the noninteracting fixed point  as \cite{Burmistrov2012,Foster2012}
\begin{equation}
\frac{d\gamma}{d \ell} = \underbrace{\bigl (x_\textsf{(1)} - x_\textsf{(2)} \bigr )}_{=0} \gamma + v  \gamma^2 + O(\gamma^3) .
\label{eq:gamma:exact}
\end{equation}
Due to Weyl symmetry $x_\textsf{(1)} = x_\textsf{(2)}$ such that the right hand side of  Eq. \eqref{eq:gamma:exact} starts from the term of the second order in $\gamma$. 
However, the coefficient $v$ is zero within the one-loop approximation(lowest order in $t$). It would be interesting to compute renormalization of $\gamma$ to the two-loop order and to find  
the coefficient $v$. Also one could try to find $v$ via the properties of pure scaling operators and their product expansions at the noninteracting spin quantum Hall transition. We leave these questions to be resolved in future works. 

\subsection{The dephasing rate}

The computation of the two-loop renormalization of the operator with two $Q$ matrices contains information on the dephasing rate. Similar to the case of standard Wigner-Dyson classes, two-loop results \eqref{eq:trw24} and \eqref{eq:trw25} (for $\sgn(\varepsilon_n \varepsilon_m)<0$) can be interpreted as the diffuson self-energy correction to the one-loop result \eqref{eq:P21:r}. After the analytic continuation to the real frequencies the real part of the diffuson self-energy at coinciding energies and zero momentum is nothing but the dephasing rate given as (see details in Refs. \cite{Burmistrov2015m,Burmistrov2016}) 
\begin{gather}
\frac{1}{\tau_\phi(E)} = \frac{3 \gamma}{2 D} \int \limits_{-\infty}^\infty d\omega\int_{q} \Bigl (2\mathcal{B}_\omega-\mathcal{F}_{\omega-E}-\mathcal{F}_{\omega+E}\Bigr ) \notag \\
\times \Re \mathcal{D}_q^R(\omega) \Im\left [ \frac{\mathcal{D}_q^{t,R}(\omega)}{\mathcal{D}_q^R(\omega)}\right ]  .
\label{eq:dephasing}
\end{gather}
Here we introduce $\mathcal{B}_\omega=1/\mathcal{F}_{\omega}=\coth(\omega/2T)$ whereas $\mathcal{D}_q^R(\omega)$ and $\mathcal{D}_q^{t,R}(\omega)$ denote retarded propagators corresponding to Matsubara propagators, cf. Eqs. \eqref{eq:prop:def:a} and \eqref{eq:prop:def:b}. We note that the expression \eqref{eq:dephasing} coincides upto numerical factor with the contribution to the dephasing rate due to exchange interaction in the case of class AI. At the Fermi level, $E=0$, Eq. \eqref{eq:dephasing} coincides with the
expression found in Ref. \cite{Bruno2005} for the class C symmetry. We stress that the result \eqref{eq:dephasing} is valid in the regime of weak disorder, $t\ll 1$, alone. At the spin quantum Hall transition one expects a power law dependence of the dephasing rate on temperature. For weak interaction, $|\gamma|\ll 1$, one should be able to express the dephasing rate exponent in terms of the generalized multifractal exponents for operators with two and four $Q$ matrices as it was done in the case of the integer quantum Hall transition \cite{Burmistrov2011}. 

In the presence of strong spin-orbit coupling the triplet diffusive modes become massive and the theory \eqref{eq:NLSM} reduces to the {\NLSM} in class D. We emphasize that in this case it is not possible to construct the Finkel'stein-type interaction term in {\NLSM}. Consequently, the renormalization group flow of $t$ remains protected against interaction. Electron-electron interaction is responsible for the dephasing rate only. It would be interesting to study the dephasing rate at the thermal quantum Hall effect (class D) using the approach of Ref. \cite{Burmistrov2011}. 

\subsection{Reduction to class A}

If the spin-orbit induced scattering is of the Ising type the triplet diffusive mode with the total spin projection 1 remains massless. This situation corresponds to the breaking of SU(2) spin symmetry down to U(1). Then the $Q$ matrix in the spin acquires the diagonal form 
\begin{equation}
Q=\begin{pmatrix}
Q_\uparrow & 0 \\
0 & Q_\downarrow 
\end{pmatrix}, 
\qquad  Q_\downarrow = - L_0 Q_\uparrow^\texttt{T} L_0 .  
\end{equation}
Using the relation between $Q_\downarrow$ and  $Q_\uparrow$, we can express the action \eqref{eq:NLSM} in terms of $Q_\uparrow$ solely. Then the action acquires the form of the Finkel'stein {\NLSM} in (spinless) class A. Now $\gamma$ plays the role of the dimensionless interaction parameter in the singlet channel. The renormalization group equations for $t$, $Z_\omega$, and $Z$ can be 
obtained from Eqs. \eqref{eq:RG:one-loop} by setting $\textsf{v}$ to 0 and reducing interaction contributions by a factor of 3. Equation for $\gamma$ can be derived from the condition that the spin susceptibility is not renormalized, $\delta Z_\omega+\delta \Gamma_t=0$. It leads to the one-loop equation for the singlet interaction amplitude 
known for class A \cite{Baranov1999b}
\begin{equation}
\frac{d\gamma}{d\ell} = -\gamma(1+\gamma) t \quad \textrm{(class A)} . 
\end{equation}
The eigenoperators can be also expressed in terms of $Q_\uparrow$ alone. 
The anomalous dimensions of these eigenoperators are given by Eq. \eqref{eq:eta:final:q} in which the interaction contribution is reduced by a factor 3 and 
$\mu_{2,1,\dots,1}$ is given by Eq. \eqref{eq:eta:final:q:2} with $\textsf{v}=0$ and $\textsf{c}_\textsf{j}=1-2\textsf{j}$.

We note that typically exchange interaction $\gamma$ in normal two-dimensional systems is positive. However, the attractive Cooper channel interaction can reverse the sign of the exchange interaction \cite{Burmistrov2012,Burmistrov2015b}. Therefore, it is natural to expect that for the class C the interaction parameter in negative and lies in the range $-1\leqslant\gamma<0$, see Eqs. \eqref{eq:deltag:1loop} and \eqref{eq:BGF:Zren}. 
After reduction to the class A the range $-1<\gamma<0$ corresponds to the repulsive short-ranged interaction in the singlet particle-hole channel whereas $\gamma=-1$ holds for the case of Coulomb interaction. 

\subsection{Summary}

To summarize we developed the theory of generalized multifractality in class C in the presence of interaction. Using the background field method we constructed the pure scaling derivativeless operators in the Finkel'stein {\NLSM} in class C. As in the standard Wigner-Dyson classes these operators in the interacting theory are straightforward generalization of corresponding operators for the noninteracting case. Employing second order perturbation theory in inverse spin conductance, we computed the anomalous dimensions of the pure scaling operators. These anomalous dimensions are affected by the interaction. Additionally, we checked that the constructed operators are indeed eigenoperators with respect to the renormalization group. Application of our results to the transition in $d=2+\epsilon$ dimensions demonstrates that interaction breaks the exact symmetry relations between generalized multifractal exponents $x_\lambda$ known in the absence of interaction.
As a byproduct of our analysis, we reproduced the results known in the literature  for the one-loop renormalization of the spin conductance, the interaction, the Finkel'stein frequency renormalization parameter, and the disorder-averaged {\DOS}. We discussed future developments and applications of our theory.

\begin{acknowledgements}
		
The authors are grateful to J. Karcher, A. Mirlin for initial collaboration on the project and for very useful discussions. One of us (I.S.B.) thanks I. Gornyi, I. Gruzberg, and H. Obuse for collaboration on related project. S.S.B. is grateful to the Karlsruhe Institute of Technology for hospitality. The research was  supported by Russian Science Foundation (grant No. 22-42-04416).  

\end{acknowledgements}

\bibliography{literature_classC}	
	
\end{document}